\documentclass[sigconf,10pt]{acmart}

\usepackage{xcolor}
\usepackage{colortbl}
\usepackage{pgfplots}
\usepackage{tikz}
\usepgfplotslibrary{groupplots}
\usepackage{bbm}
\usepackage{float}
\usepackage{subcaption}
\usepackage{multirow}
\usepackage{cancel}

\usepackage{microtype}
\usepackage{graphicx}
\usepackage{booktabs} 
\usepackage{tabularx}

\usepackage{hyperref}

\DeclareFontFamily{U}{mathc}{}
\DeclareFontShape{U}{mathc}{m}{it}%
{<->s*[1.03] mathc10}{}

\newcommand{\statreport}[1]{\ensuremath{\left[#1\right]}}

\AtBeginDocument{%
  \providecommand\BibTeX{{%
    \normalfont B\kern-0.5em{\scshape i\kern-0.25em b}\kern-0.8em\TeX}}}

\setcopyright{acmcopyright}
\copyrightyear{2018}
\acmYear{2018}
\acmDOI{XXXXXXX.XXXXXXX}

\acmConference[Conference acronym 'XX]{Make sure to enter the correct
  conference title from your rights confirmation emai}{June 03--05,
  2018}{Woodstock, NY}
\acmPrice{15.00}
\acmISBN{978-1-4503-XXXX-X/18/06}

\begin{document}

\title{LightningNet: Distributed Graph-based Cellular \\Network Performance Forecasting for the Edge}

\author{Konstantinos Zacharopoulos}
\affiliation{%
  \institution{Technical University of Crete}
  \country{Greece}}
\email{constantine.zach97@gmail.com}

\author{Georgios Koutroumpas}
\affiliation{%
  \institution{Telef\'{o}nica Research}
  \country{Spain}}
\email{george.koutroumpas@telefonica.com}

\author{Ioannis Arapakis}
\affiliation{%
  \institution{Telef\'{o}nica Research}
  \country{Spain}}
\email{ioannis.arapakis@telefonica.com}

\author{Konstantinos Georgopoulos}
\affiliation{%
  \institution{Technical University of Crete}
  \country{Greece}}
\email{kgeorgopoulos@tuc.gr}

\author{Javad Khangosstar}
\affiliation{%
  \institution{Virginmedia O2 UK}
  \country{United Kingdom}}
\email{javad.khangosstar@virginmediao2.co.uk}

\author{Sotiris Ioannidis}
\affiliation{%
  \institution{Technical University of Crete}
  \country{Greece}}
\email{sotiris@ece.tuc.gr}

\renewcommand{\shortauthors}{Trovato and Tobin, et al.}

\begin{abstract}
The cellular network plays a pivotal role in providing Internet access, since it is the only global-scale infrastructure with ubiquitous mobility support. To manage and maintain large-scale networks, mobile network operators require timely information, or even accurate performance forecasts. In this paper, we propose LightningNet, a lightweight and distributed graph-based framework for forecasting cellular network performance, which can capture spatio-temporal dependencies that arise in the network traffic. LightningNet achieves a steady performance increase over state-of-the-art forecasting techniques, while maintaining a similar resource usage profile. Our architecture ideology also excels in the respect that it is specifically designed to support IoT and edge devices, giving us an even greater step ahead of the current state-of-the-art, as indicated by our performance 
experiments with NVIDIA Jetson.
\end{abstract}

\received{20 February 2007}
\received[revised]{12 March 2009}
\received[accepted]{5 June 2009}

\maketitle

\section{Introduction}
\label{sec:introduction}

The rapid, and often unplanned, urbanization has been a critical driver behind many social and critical infrastructure changes. The significant mobility and sustainability challenges that these changes bring can only be further aggravated, as this unprecedented transition from rural to urban areas continues. For instance, recent urbanization practices, along with the widespread adoption of IoT-based technologies and mobile devices by the market~\cite{li2015}, have created a greater need for better telecommunication services and faster Internet connectivity. Furthermore, billions of mobile users are accessing the Internet as they move. For their Internet access, the cellular network plays a pivotal role, since it is the only global-scale infrastructure with ubiquitous mobility support. 

However, the performance of a cellular network is dynamic; it changes over time, it is prone to hardware and software failures, and it is affected by regular (e.g., holiday events, seasonal changes) and irregular events (e.g., traffic, weather conditions, human behaviour)~\cite{Mahimkar2013}. As a result, mobile network operators (MNOs) make large investments to accurately monitor and understand performance dynamics, as they are key to an effective resource allocation, planning~\cite{mishra2004}, and optimization~\cite{Laiho2001}. For the most part, operators rely on a set of key performance indicators (KPIs) that are - in principle - hardware measurements correlated with performance. Key performance indicators offer insights about each cellular sector over a certain time window~\cite{Kreher2015} and can be broadly classified into three main categories: (1) signaling and coverage monitoring; (2) voice-related measurements; (3) data-related metrics. To this end, MNOs consider combinations of multiple network measurements, which reflect the overall performance of individual sectors~\cite{Kaiping2008, Agarwal2008, Nika2014, Leontiadis2017} and use them in combination with AI-based network optimization techniques. Therefore, central to the mobility management of a cellular network is a high-value task: network performance forecasting, hereinafter referred to as \textit{hot spot forecasting}. 

The methodology to combine KPIs into a single metric and the threshold to determine hot spots have been established over the years by vendors and the industry based on domain knowledge, service level agreements, and controlled experiments~\cite{Laiho2001}. Being able to forecast hot spots offers a number of advantages, e.g., (1) investment plans can be finalized in advance, since anticipating future demands allows operators to optimize CapEx spending~\cite{mishra2004}; (2) it permits operators to proactively troubleshoot their network~\cite{Winstein2013}; (3) it makes possible a dynamic resource allocation as an input to self-organizing networks ~\cite{Waldhauser2012}.

Considering the dimensionality of said problem space, efficient optimization strategies are often implemented in a cloud environment using sufficiently powerful computational hardware. Some prior attempts have explored such centralized forecasting solutions~\cite{Serra2017, Dalgkitsis2018}. However, these solutions rely for the most part on conventional Machine Learning (ML) algorithms with limited forecasting capabilities~\cite{Serra2017}, that do not support a joint modelling of spatio-temporal patterns, and often take a na\"{i}ve approach of central data collection at the cloud which is not scalable to the number of edge devices, depletes their residual energy, and may not be even possible when connectivity is only periodically available.

Based on the need to take some burden off the cloud, vendors of reconfigurable hardware have started to provide Multi-Processor Systems-on-Chip (MPSoCs) that integrate multiple general-purpose CPUs, GPUs, hardwired accelerators and programmable engines in the form of embedded computing boards, e.g., Jetson series, Flo Edge, Brainy pi, etc. Additionally, the nature of edge AI networks, where edge devices are situated in dynamic and unpredictable environments, has motivated further the adaptive allocation of computational tasks to the edge and fog devices. Our paper, which is premised on this edge-centric view, proposes a lightweight approach that can address the problem of hot spot forecasting in a distributed manner, by deploying at the edge novel graph-based algorithms for building efficiently compact data and model summaries. Such summaries reduce the size of the data processed per edge device, in order to refine a local model, and can evolve with incoming data to support real-time decision-making at the edge. Moreover, only summaries of local models and data are communicated in-network, making (1) efficient use of the available bandwidth and edge resources (e.g., battery), and (2) supporting privacy. Another advantage offered by edge computing, is the reduction of latency due to communicating information from a single device across the network to a centralized computing system or cloud server. Beyond these, our key contributions to the hot spot forecasting problem are the following:

\begin{itemize}
\item We consider a large-scale dataset of hourly measurements over a period of two months, comprising tens of thousands of sectors, collected for a whole country by a major european mobile operator with more than 25 million subscribers.
\item We propose an approach to featurising the cellular network data and presenting it to the  Graph Neural Networks (GNNs). Our lightweight data and model summarization minimizes network communication and supports efficient model refinement in resource-constrained, federated learning scenarios.
\item We provide a flexible and nimble forecasting framework that allows using KPIs to predict future hot spots. We compare and contrast our approach against SoA centralised solutions and demonstrate sizeable improvements.
\item We evaluate the performance for different forecasting horizons and amount of historical information. Among others, we show that the proposed forecasting framework achieves a stable performance (in terms of precision) even when targeting an horizon of 48 hours, while there is no clear pattern between the alterations of historical information in the input and the model's recall.
\item We conduct a meta-analysis on the use of sub-graph split, their transferability and the benefits of applying a hierarchical architecture.
\item Finally, we drop the assumption that AI models and systems need to be centrally trained and demonstrate the integration of hardware accelerators (MPSoC/FPGA) with edge devices, which can support real- time decision-making at the edge for radio access networks. 
\end{itemize}

\section{Related Work}
\label{sec:related_work}

Traffic and cellular performance forecasting have both been hot topics of research in the recent years. In this section, we review the most important works that informed the proposed model architecture and discuss their existing limitations, which we address. Being a combination of Recurrent Neural Networks (RNNs) and Graph Neural Networks (GNNs), our model directly inherits the benefits those works have been credited with.

\subsection{Multivariate Time Series Forecasting}
\label{ssec:mtsf}

Traffic forecasting belongs to multivariate time series analysis and has been studied for decades. Machine learning methods capable of leaning complex, non-linear relationships, such as support vector regression (SVR) methods~\cite{Smola2004ATO}, tend to outperform their statistical counterparts like the Autoregressive Integrated Moving Average (ARIMA)~\cite{Hamed1995} and its variants, with respect to handling complex traffic forecasting problems~\cite{MA2015187}. In recent years, the introduction of high-capacity function approximators (i.e. deep neural networks), along with effective algorithms for training them, has shown excellent results compared to conventional ML methods, and has allowed learning from big data collections in an automatic manner and for a number of domains, e.g., traffic forecasting~\cite{POLSON20171}. Many NN-based models~\cite{Chung2015, Luo2018, Rangapuram2018, Lai2018, Cao2018, Tang2020JointMO, Salinas2020, Can2020}, including RNNs~\cite{Hopfield1982, Rumelhart1986, Jordan1997}, have been proposed to model temporal dynamics in time series data. In particular, RNNs employ artificial memory to maintain information patterns through cyclical runs on the input data. Due to that memory component, RNN algorithms can achieve very high accuracy, but are also invariant to the sequence length of the time series. Some examples of popular RNN architectures are Gated Recurrent Units (GRUs)~\cite{ChungGCB14,ChoMGBSB14}  and Long Short Term Memory~\cite{Hochreiter1997}. 

RNNs have been widely adopted as a component of a traffic forecasting model to predict traffic speed~\cite{MA2015187}, travel time~\cite{Duan2016}, and traffic flow~\cite{Zhao2017, SAEs, RNN_traffic}. Within this body of research, one of the few works that addressed the hot spot forecasting problem and provided an in-depth analysis of the dynamics of cellular network congestion points, is by~\citet{Serra2017}, where they present the problem as a multivariate time series forecasting task and consider the temporal patterns in the network traffic. The authors, who focused mainly on temporal regularities, demonstrated that the proposed tree-based models can introduce significant performance improvements over the baselines, but these improvements become less impactful for long-range horizons. Moreover, their forecasting method does not account for the topology of the network, nor exploits the learning of relations between nodes in a graph and the rules for composing them. Another recent work is that of~\citet{LSTM_HOTSPOT}, which addresses the problem of hot spot forecasting using an LSTM based model trained on data of a Chinese city. Although the properties of the LSTM make it possible to deliver competitive forecasting performance for longer prediction horizons, their predictive approach does not consider network topology or the structural features governing its composition.

\subsection{Convolutional Neural Networks}
\label{ssec:cnn}

To model the spatial relationships observed in traffic networks, some prior works~\cite{Ma2017LearningTA, Jin2018, Yu2017SpatiotemporalRC} have considered Convolution Neural Networks (CNNs), also in conjunction with LSTMs~\cite{Zhao_2021,app12178714} for traffic flow prediction, and have been shown to work effectively for short-term forecasting tasks~\cite{DiYANG20192018EDP7330}. Convolution Neural Networks employ learnable sets of filters, known as kernels, to perform convolutions and extract spatial features from 2D spatial-temporal data. Due to the attributes of convolution, these learned patterns are invariant to their location, making their detection optimal in multidimensional data. 

One of the existing challenges is converting traffic network structures to images and using CNNs to learn spatial features. Inevitably, these converted images introduce a certain amount of noise, resulting in spurious spatial relationships captured by CNNs. Due to the inefficacy of conventional CNN methods in dealing with the topological structure and physical attributes of traffic networks (or cellular, in our case), certain works have proposed to learn the traffic network as a graph and adopt the graph-based convolution operator to extract features from the graph-structured traffic network, which we review next.

\subsection{Graph Neural Networks} 
\label{ssec:graph-nn}

Graph Neural Networks are a family of algorithms used to efficiently operate and generalize over environments represented as graphs. Traffic networks have been analyzed as graphs for dynamic shortest path discovery~\cite{Sun2017}, traffic congestion analysis~\cite{SUN2014496}, and dynamic traffic assignment~\cite{Kalafatas2007}. These works build directly upon previous research in geometric representation learning and, in particular, on the Graph Network framework~\cite{Battaglia2018} and the encode-process-decode~\cite{Hamrick2018} paradigm. This allows them to align better with the iterative nature of traffic computations, as well as path-finding algorithms.

Recently, several studies ~\cite{Henaff2015,Kipf2016} attempted to generalize neural networks to work on arbitrarily structured graphs, by introducing Graph Convolutional Networks (GCNs). The GCNs rely on an adjacency or Laplacian matrix to represent the structure of a graph. In the context of traffic forecasting, traffic network can be represented as graphs consisting of nodes and edges, and thus, several GCN-based models, including the spectral graph convolution~\cite{Yu_2018} and the diffusion graph convolution~\cite{Atwood2016, Yaguang2017}, have been shown to be suitable for network-wide forecasting tasks. Additionally, to detect the spatio-temporal dynamics of traffic data flows, GCNs have been combined with RNNs. In~\cite{Zhao_2020}, GCNs have been fuzzed with GRU and LSTM components, to perform traffic control. There, the combination with GRU outperformed the LTSM counterpart, while also outperforming GCN, GRU, ARIMA and SVR algorithms. Another example of work using a GCN-LSTM architecture~\cite{https://doi.org/10.48550/arxiv.1802.07007} has been shown to handle effectively network scale data, similar to our use case.

All the above works have addressed the problem of traffic forecasting. To the best of our knowledge, this is the first work that proposes a distributed, graph-based architecture (consisting of a GCN module and an RNN module), capable of accurately predicting hot spots across a vast network, by operating on different subgraphs. Our lightweight ML framework, which manages to extract both spatial and temporal patterns, is not only suitable for edge-processing scenarios but also creates new research possibilities for graph-mining, such as determining the optimal subgraph combination that provides the best performance for the cellular network performance forecasting task.
\begin{figure}[!t]
    \centering
    \includegraphics[width=0.48\textwidth]{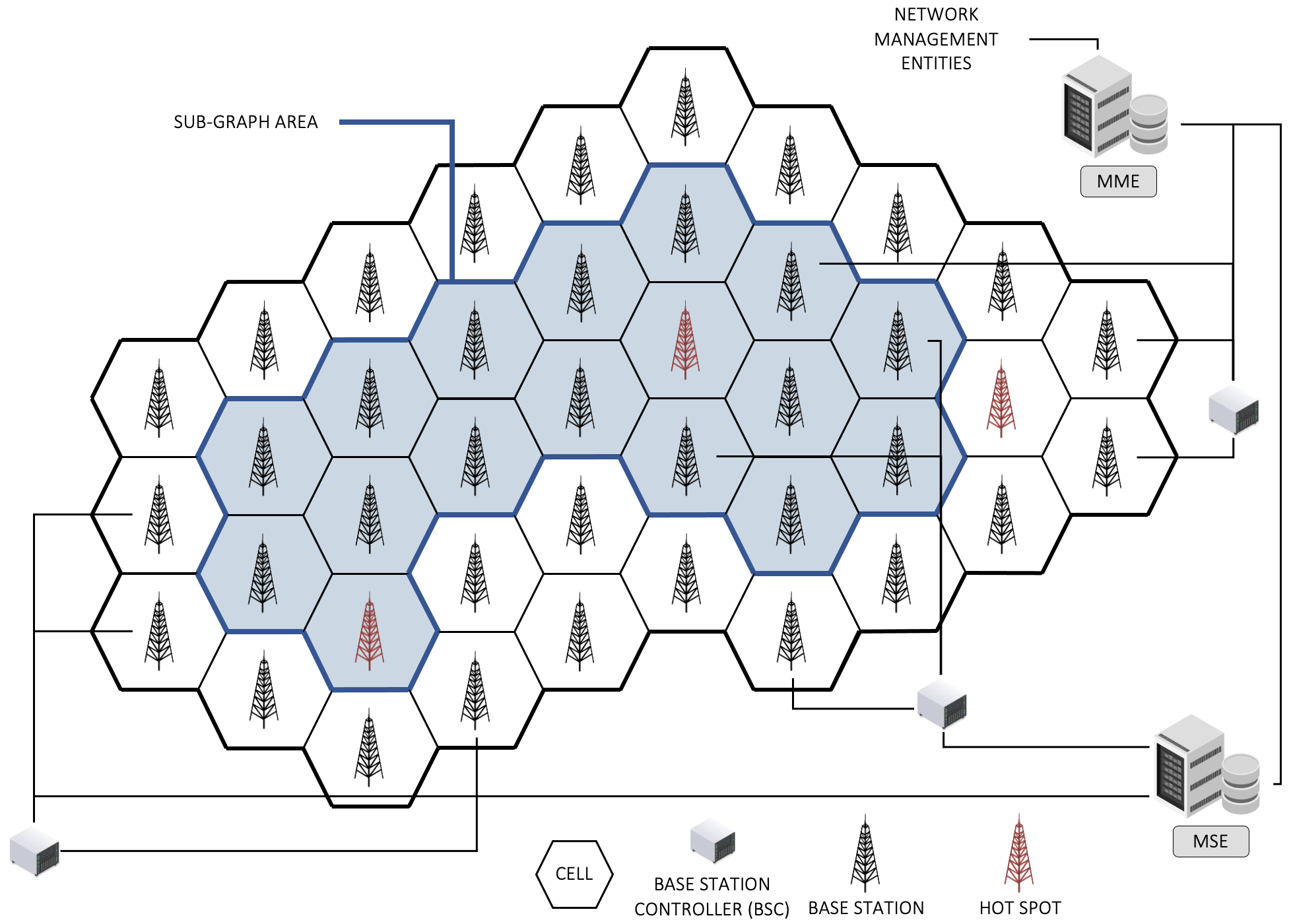}
    \caption{High-level architecture of the measurement and forecasting infrastructure integrated in the cellular network.} 
    \vspace{-0.3cm}
    \label{fig:high-level-architecture-1}
\end{figure}

\section{Methodology}
\label{sec:methodology}

\subsection{Dataset}
\label{ssec:dataset}

\subsubsection{Measurement Infrastructure}
\label{sssec:measurement_infrastructure}

Here we describe the measurement infrastructure we leverage for collecting network data from a large commercial MNO in the UK (with more than 25\% market share in the UK in 2019). The cellular network we study supports 2G, 3G, 4G and 5G mobile communication technologies. Such a network can be simplified to consist of three main domains: (1) the cellular device (in our case, the smartphone used as primary device by end-users), (2) the Radio Access Network (RAN) and (3) the Core Network (CN). For our analyses, we focus only on domains (2) and (3). Figure ~\ref{fig:high-level-architecture-1} sketches the high-level architecture of a 4G mobile network such as the one used in our work.

\noindent\underline{RAN:} The radio access consists of hundreds of thousands of components (e.g., sectors, towers, and controllers). Those compose the mobile network ``last frontier'', and bridge the users' devices with the core network, which enables access to voice calls and the Internet. To support troubleshooting and optimization, the vendors provide monitoring platforms to (passively) collect KPIs from radio access network elements and backbone links. To simplify the monitoring complexity, KPIs normally are aggregated over time (with periodicity varying from minutes to hours) and users.

\noindent\underline{Cell Sites:} Cell sites (also called cell towers) are the sites where antennas and equipment of the RAN are placed. Every cell site hosts one or multiple antennas for one or more technologies (i.e., 2G, 3G, 4G), and includes multiple cells and sectors. For every cell site we have detailed information including location, radio technologies available, number of cells and radio sectors. We collect aggregated KPI for every radio sectors (e.g., radio load, average throughput).

\subsubsection{Data Feeds}
\label{sssec:data_feeds}
From our measurement infrastructure, we capture various data feeds from the mobile network that we describe next. Our dataset spans across July-August of 2021 and consists of 15,118,080 hourly observations for several thousands of cells that are aggregated at postcode level or larger granularity. Due to the proprietary and confidential nature of our dataset, we are constrained to the amount of details we can disclose. We further note that, previous studies have also confronted such issues, rendering the implementation of baseline methods a challenging task (see Sec.~\ref{sssec:baselines}). 

\noindent\underline{Radio Network Topology:} To account for potential structural changes in the radio access network (e.g., new site deployments), we rely on a daily snapshot of the network topology. This includes metadata (location and configuration) and the status (active/inactive) of each cell tower.

\noindent\underline{Radio Network Performance:} We rely on a commercial solution the MNO deploys to collect the radio network performance dataset. This dataset includes various KPI, including average cell throughout, average user throughput, average percentage of resources occupied, average number of users, total volume of data traffic uplink/downlink and total volume of conversational voice traffic. 

\subsubsection{Network KPIs}
\label{sssec:network_kpis}

With respect to cellular network performance, there is no definitive list of KPIs. However, through a joint effort, vendors and operators alike have identified a set of KPIs whose importance is acknowledged by multiple parties~\cite{Holma2007, Agarwal2008, ting2008, Kreher2015}. We categorize KPIs into five different groups~\cite{Leontiadis2017}:

\begin{itemize}
\item \textbf{Signaling}: These KPIs are mostly related to faults such as failure to establish a Radio Access Bearer (RAB), or Radio Resource Control (RRC), or the fact that the sector cannot efficiently reach the radio controllers.
\item \textbf{Voice}: These KPIs capture failure to establish or maintain a voice call.
\item \textbf{Data availability}: These metrics reflect the availability of high-speed data channels (e.g., HSDPA/HSUPA/LTE) at any given time or the number of data-active connected devices.
\item \textbf{Data Congestion}: These metrics capture the fact that the capacity was reached. For instance, the average number of users queuing to get an HSDPA/HSUPA/LTE channel, or the number of times a data connection had to be dropped to make room for a voice call. Furthermore, they indicate the percentage of time that the radio was active transmitting data.
\item \textbf{Radio}: Radio KPIs are about interference, power statistics, wireless noise, signal conditions, etc.
\end{itemize}

\subsubsection{Deriving the Hot Spot Score}
\label{sssec:ground_truth}

The primary use of KPIs is to enable operators to monitor the health of the network and to quickly identify bottlenecks. Therefore, it is natural to use KPIs as a way to flag network conditions that deteriorate below an established performance limit~\cite{Agarwal2008, Kaiping2008, Alam2013}. As a result, for each KPI, a threshold has been set. Such thresholding mechanisms are widely used in the industry, and allow network planners and radio resource operators to focus their attention where it is required. Default threshold values are proposed by the equipment vendors, while operators further fine-tune them based on their experience, objectives, and domain knowledge. Hence, significant investments are made through drive tests, controlled experiments, and A/B testing, to identify performance bottlenecks, and how these relate to the KPIs and thresholds~\cite{Alam2013, Ericsson2014, mishra2004, Huawei2006, ting2008}.

While establishing thresholds for each KPI enables a fine-grained vision of specific problems, each network element is associated with hundreds of metrics. It is then desirable to consolidate measurements into a single performance index so to (1) quantify the “health” of each network element; (2) easily assess the whole network status and trends; (3) narrow down on sites that need attention. Such an example is the hot spot score~\cite{Agarwal2008, Kaiping2008, Nika2014}, which represents how ``hot'' or problematic a given sector is. It is a weighted combination of thresholded KPIs related to signaling, voice, data availability and data congestion, such as the one proposed by~\citet{Leontiadis2017}:

\begin{equation}
    s_{b} = S(\mathbf{k}_{b}, \mathbf{w}, \mathbf{t}) = \sum_{i=1}^{n}w_{i} \cdot H(k_{b,i} - t_{i})
\end{equation}

where $b$ is the sector under study generating $n$ KPIs collected in the vector $\mathbf{k}_{b}$. The KPIs are associated to weight $\mathbf{w}$ and threshold $\mathbf{t}$ vectors, while $H(\cdot$) is the Heaviside step function which outputs 1 when the KPI value $k_{b,i}$ reaches the corresponding threshold $t_{i}$ and 0 otherwise. The higher the score $s_{b}$, the more ``hot'' the sector $b$ is. In summary, a hot spot score is a linear combination of the weights associated to KPIs that trigger.

In the pursuit of ecological validity, we draw upon MNO data characterised by hotspot distributions that mirror real-world conditions. More specifically, in our analysis, we use as ground-truth a very similar metric computed using a \textit{proprietary} hot spot formula that cannot be disclosed. We further note that our data suffers from class imbalance. Within the months of July-August, we observe $\leqq 0.25\%$ of hot spot cases. Despite the relatively low percentage of hotspots, these events can have a disproportionate effect on network performance, thereby incurring direct implications for investment planning and CapEx optimization. Moreover, in such scenarios, classifiers are biased to learn and predict the majority class, leading to superficial accuracy values (a known issue in machine learning), which we address by considering appropriate performance metrics.

\begin{table}[!t]
\begin{center}
\resizebox{0.99\linewidth}{!}{
\begin{tabular}{ccccccc}
\toprule
\\
& \multicolumn{1}{c}{\texttt{Forward}} & \multicolumn{1}{c}{\texttt{Backward}} & \multicolumn{1}{c}{\texttt{Stepwise}} & \multicolumn{1}{c}{\texttt{CorMap}} & \multicolumn{1}{c}{\texttt{LASSO}} & \multicolumn{1}{c}{\texttt{RIDGE}}\cr
\\
\midrule
\\
Model Precision(\%) & 0.49 & 0.47 & 0.52 & 0.43 & \textbf{0.55} & 0.51\\
Run-time(sec) & 4,027 & 3,918 & 9,878 & 193 & 2,519 & 2,384 \\
\\
\bottomrule
\end{tabular}
}
\caption{Feature Selection algorithm performance.}
\label{featResults}
\end{center}
\end{table}

\begin{table}[!t]
\begin{center}
\resizebox{0.99\linewidth}{!}{
\begin{tabular}{cccccccc}
\toprule
\\
& \multicolumn{1}{c}{\texttt{Zero Filling}} & \multicolumn{1}{c}{\texttt{Mean}} & \multicolumn{1}{c}{\texttt{Median}} & \multicolumn{1}{c}{\texttt{Most Frequent}} & \multicolumn{1}{c}{\texttt{Hot Deck}} & \multicolumn{1}{c}{\texttt{Cold Deck}} & \multicolumn{1}{c}{\texttt{KNN}}\cr
\\
\midrule
\\
Imputation Accuracy(\%) & 0.39 & \textbf{0.61} & 0.57 & 0.48 & 0.60 & 0.58 & \textbf{0.64}\\
Run-time(sec) & 5,732 & \textbf{10,621} & 9,105 & 9,009 & 16,318 & 15,758 & 25,122\\
\\
\bottomrule
\end{tabular}
}
\caption{Imputation method performance.}
\label{imputeTableResults}
\end{center}
\end{table}

\subsubsection{Data Preprocessing}
\label{sssec:data_preprocessing}

Our initial dataset consists of 77 KPIs, which span across different categories of data feeds (see Sec.~\ref{sssec:network_kpis}). These KPI measurements may not always be registered for every sector, hour and indicator due to various reasons (i.e. broken/inoperable sensors). To prevent data corruption during training, we treat item non-response cases by applying known imputation algorithms (e.g., scikitlearn implementations\footnote{\url{https://scikit-learn.org/stable/modules/impute.html}} as shown in Table ~\ref{imputeTableResults}. The cases of unit non-response are treated by copying the sample from the previous recorded time-step. In addition, in order to select the most discriminatory KPIs for our hot spot forecasting task we perform a combination of feature selection techniques. First, eliminate those KPIs with near zero-variance. Then we consider various wrapper, filter and embedded feature selection algorithms as shown in Table \ref{featResults}. Our feature selection step removes redundant features and reduces our KPIs down to 35, which introduces significant efficiency improvements to our model training
pipeline.

\subsection{Hot spot Forecasting}
\label{ssec:problem}

\subsubsection{Preliminaries}
\label{sssec:preliminaries}

While different versions of GCNs have been proposed in the literature, they all share some basic principles. The assumption of GCNs is that the embedding of a node must result from the aggregation of itself and its neighbours. More specifically, GCNs aim to learn a function of node features on  graph topological data. It is  defined as a function over the undirected graph $G=(V,E)$ where $V$ denotes the vertices or nodes $v_{i} \in V$ and $E$ denotes the edges $(v_{i},v_{j} \in E)$. The nodes $V$ of the graph represent the cells and the edges $E$ represent their physical distance, indicating whether or not two antennas are considered neighbours. To define the boundaries of such neighbourhoods, we use the geolocations of the cells and calculate the pairwise Geodesic Distance between them by finding the angle between the two cells, multiplied by the circumference of the earth:

\begin{equation}
    \begin{split}
        angle &= \arccos(cell_{i} * cell_{j}) \\
        dist &= angle * pi * R
    \end{split}
\end{equation}

where ${cell}_{i}, {cell}_{j}$ are the latitude and longitude values of the two cells and $R$ is the radius of the earth. This produces a strictly upper triangular distance matrix ${D} \in \mathbb{R}^{|M|\times |M|}$, where $M$ is the total number of cells in the network. To calculate the adjacency matrix, we consider different distance thresholds (measured in km) $t \in \{0.0,0.1,0.5,1.0,1.5,3.0\}$. If $dist \leqq t$, cells ${cell}_{i}, {cell}_{j}$ are considered neighbours. This process provide us with a strictly upper triangular adjacency matrix ${A} \in \mathbb{R}^{|M|\times |M|}$. 

\subsubsection{Baseline Models}
\label{sssec:baselines}

\hfill \break
\noindent\underline{LSTM:}
Long Shot-Term Memory~\cite{Hochreiter1997} is a kind of RNN designed to be used in language models. The LSTM architecture is well known for its ability to capture long-term dependencies, while also being able to prevent the so called vanishing/exploding gradient problem that RNNs typically suffer from. An LSTM model contains an internal self-looping module, allowing for data flow both forward and backward in the network. This enables the network to keep memory of old information, making it a strong candidate for time series tasks. 

LSTM-based models have been successfully applied for the task of hot spot prediction~\cite{LSTM_HOTSPOT}. In this work, the authors used data by a large Chinese telecommunication service provider, which they divide into feature and target sets. Their feature set includes KPIs for over 420 cells from a northern city of China. The proposed model used for the hot spot prediction is a LSTM unit, using 24 hours of memory buffer, followed by a fully connected layer. We implement this LSTM model and compare it against our approach. To keep our analysis consistent with prior work, we test the LSTM performance with 12, 24, 36 and 48 hours of memory buffer, as well as for forecasting horizons of 12, 24 and 48 hours.

\noindent\underline{GCN:}
As our second baseline, we consider the most commonly used Spectral Method amongst the Graph Neural Network family, the Graph Convolutional Network ~\cite{Kipf2016}. The GCN provides a computationally effective way for node classification through graph convolution, which enables us to detect spatial dependencies within the cellular network.

Our data consists of sequential measurements of multiple related variables over time i.e. it's a collection of multivariate time series consisting of the KPIs for each cell. Therefore, we expect that, in each time step, the state of a cell will contribute to the accurate forecasting the future states of that cell and, to this end, we make use of the self-connections in the graph which can be enforced through the mechanics of the GCNs by adding the identity matrix $I$ to the adjacency matrix $A$:

\begin{equation}
\label{eqn:adj_mat}
     \widetilde{A} = A + I
\end{equation}

Moreover, the GCNs use the symmetric normalization of the Laplacian $L$
\begin{equation}
\label{eqn:normalization}
     L_{norm} = D^{\frac{-1}{2}} * L * D^{\frac{-1}{2}} = I - D^{\frac{-1}{2}} * A * D^{\frac{-1}{2}}
\end{equation}
and a re-normalization trick in order to tackle the exploding/vanishing gradient issues

\begin{equation}
\label{eqn:renormalization}
     I + D^{-\frac{1}{2}} * A * D^{-\frac{1}{2}} \rightarrow \widetilde{D}^{-\frac{1}{2}} * \widetilde{A} * \widetilde{D}^{-\frac{1}{2}}
\end{equation}

where $\widetilde{D}_{ij}$ is the degree matrix of the graph $G$, created through row-wise summation of its adjacency matrix 

\begin{equation}
\label{eqn:deg_mat}
    \widetilde{D}_{ij} = \sum_{j=1}^M(\widetilde{A}_{ij})
\end{equation}

Considering a feature matrix  $Z \in \mathbb{R}^{|V|\times d}$, where $d$ is the dimension of the original features, the update rule for each GCN layer can be described through the following equation

\begin{equation}
\label{eqn:update_rule}
     H^{(l+1)} = \sigma({\widetilde{D}}^{-\frac{1}{2}}*{\widetilde{A}}*{\widetilde{D}}^{-\frac{1}{2}}*{H}^{(l)}*{W}^{(l)})
\end{equation}

where W represents the trainable weight matrix.

\noindent\underline{ARIMA:}
As an additional baseline, we also consider the ARIMA algorithm. However, our time series data exhibits auto-correlation, while the behavior of a cellular network is often impacted by environmental and seasonal phenomena (i.e. increased tourism in the summer, limited mobile user movements when it rains). Such properties are dependent on the time of the observation, rendering it non-stationary, and violate the models' assumptions. Though there are ways to remove seasonality by applying differencing steps, in our case this process leads to significant information loss, rendering the ARIMA model ineffective in our problem setting.

\begin{figure}[!t]
    \centering
    \includegraphics[width=0.8\linewidth]{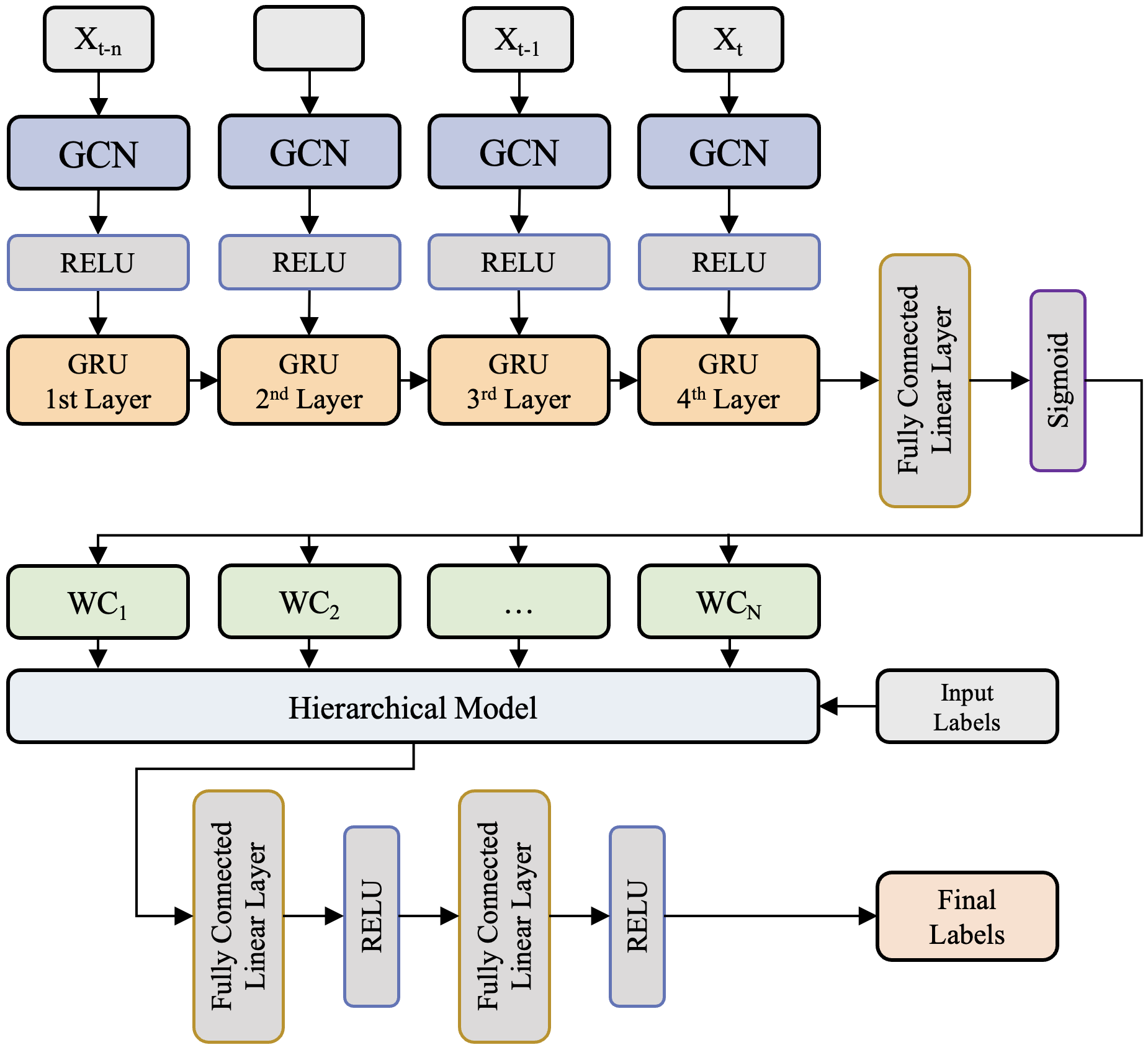}
    \caption{LightningNet Architecture. The design of the sub-classifier (SC) is depicted on the uppermost part of the diagram, whilst the lowermost part depicts the Hierarchical Model (HM) structure.} 
    \vspace{-0.3cm}
    \label{fig:model-architecture}
\end{figure}

\subsubsection{LightningNet Architecture}
\label{sssec:model_architecture}
The proposed model architecture consists of a GNN and an RNN component (Figure~\ref{fig:model-architecture}). The GNN component is a Graph Convolution layer purposed to capture spatial dependencies between neighbouring cells. 
As discussed in Sec.~\ref{sssec:preliminaries}, the adjacency matrix is defined based on the geodesic distance between the antenna cells. Therefore, smaller distance threshold values result in less populated adjacency matrices. Through experimentation, we observe for our dataset, that a highly populated adjacency matrix is associated with significant performance drops and exponentially larger training cycles. This is expected, considering that most of the cellular network infrastructure is situated in urban areas, meaning high antenna density per square kilometer. These performance drops can be explained through equations  ~\ref{eqn:deg_mat} and ~\ref{eqn:update_rule}, in conjunction with the fact that our dataset is characterized by high class imbalance. Increasing the number of antennas that will be included in the aggregation function for the update rule of the GCN results in high value corrosion. Consequently, it is necessary to fine-tune the distance threshold at low enough values in order to optimize the GCN layer.

The output of this layer is a representation of the combined feature vectors from all neighbouring cells, which is subsequently fed as input to the first layer of the RNN component. The RNN captures any temporal dependencies observed at the cell-level by using a memory buffer that stores historical information. It is a series of Gated-Recurrent-Unit (GRU) layers, the number of which may vary depending on the target prediction horizon and the temporal patterns. For each time step, the aggregated feature vector of all neighbouring cells is passed into a GRU layer and then to a fully connected layer, which then produces the one-shot prediction for the target variable i.e. the probability of becoming a hot spot in the target prediction horizon.

\begin{figure}[!t]
    \centering
    \includegraphics[width=0.76\linewidth]{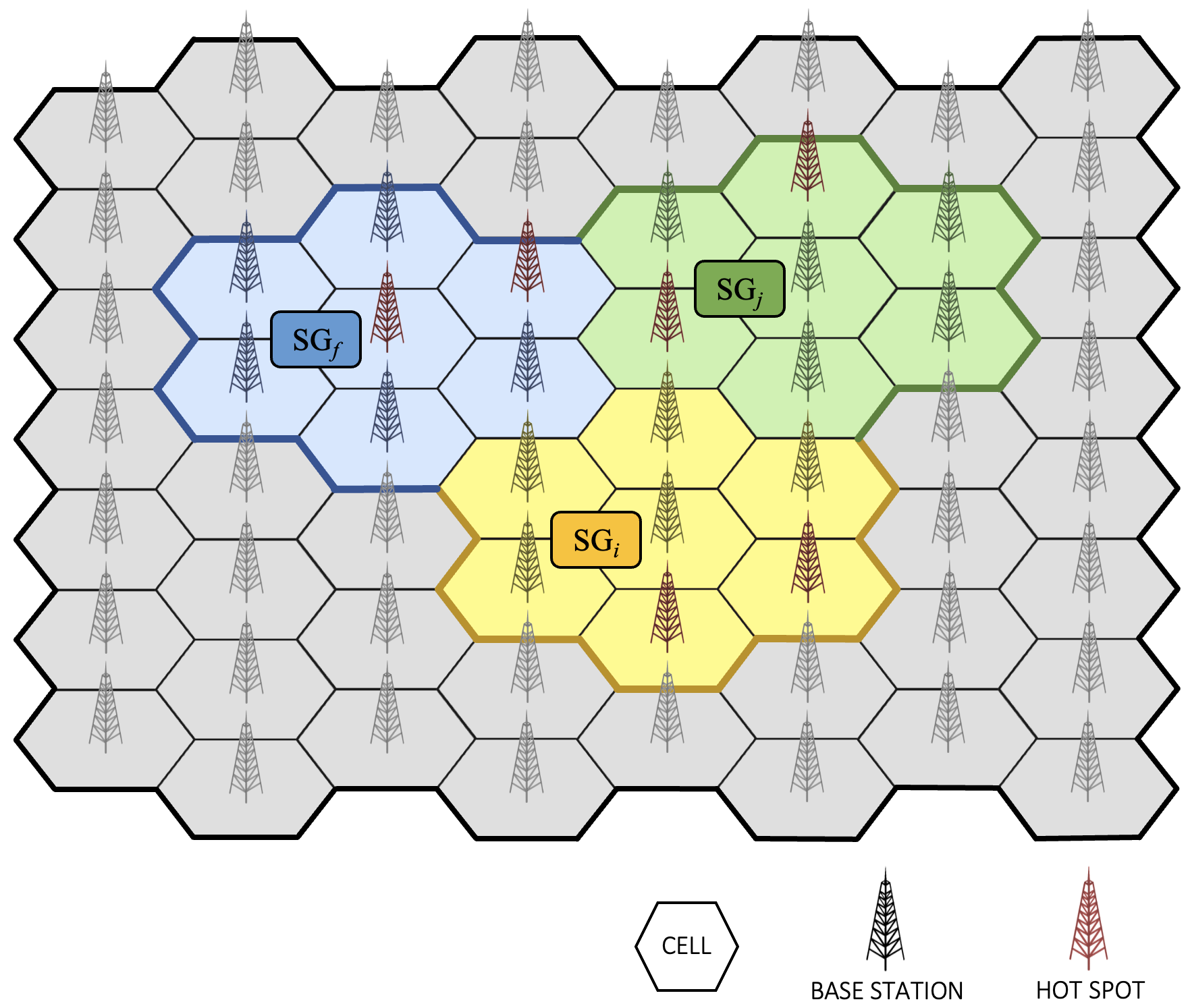}
    \caption{Graph partition into $k$ equally sized sub-graphs. The sub-graphs with the highest concentration are used to train the LightningNet model, which then can be applied on any given sub-graph in the system.} 
    \vspace{-0.3cm}
    \label{fig:hm}
\end{figure}

\subsubsection{Graph Partitioning}
\label{sssec:graph_partitioning}
Taking into account that we are dealing with a multivariate time-series with samples from several thousand antenna cells, the process of down-sampling is not a straightforward one. To reduce the volume of the processed data and produce compact model summaries, suitable for edge devices, we introduce a graph partitioning step that splits our original graph to $k$ equally-sized sub-graphs, which we use for subsequent training and testing\footnote{We preserve the chronological order of the data and derive well-separated splits to mitigate ground-truth leaks.}. This data segregation significantly reduces the system requirements (see Sec.~\ref{sec:experiments} and ~\ref{ssec:computational_overhead}) for the training and inference step, thus making it compatible with System-On-Chip devices that have limited resources. This is crucial for enabling the deployment of the proposed prognostic algorithm at the edge.

Following the graph-split, we select $k$ sub-graphs (denoted as SG$_{1-k}$) with the highest concentration of hot spot cases as shown in Figure ~\ref{fig:hm}. We then train $k$ instances of our model (i.e. $k$ ``sub-classifiers'' referred to as SC$_{1-k}$) per such sub-graph. The evaluation of the sub-classifiers is performed not only on the sub-graph they were trained upon, but also on all other sub-graphs, as part of our cross-evaluation methodology that allows us to determine the transferability of our forecasting method. High transferability is important since, once the model instances have been trained and fine-tuned, they need to perform well on any given sub-graph. Furthermore, this step allows us to significantly reduce training time and memory usage, improve on the targeted performance metrics, and validate our models on specific locations, when necessary.

As a side contribution, we investigate the correlation between sub-graph similarity as proposed in ~\cite{Koutra2011AlgorithmsFG} and model transferability. We define graph similarity $sim$ by calculating the Laplacian eigenvalues of each sub-graph's adjacency matrix. For each sub-graph, we find the smallest $k$ such that the sum of the $k$ largest eigenvalues constitutes at least 90\% of the sum of all of the eigenvalues. Then, if the values of $k$ are different between the two sub-graphs, we use the smaller one. Finally, the sum of the squared differences between the largest $k$ eigenvalues between the sub-graphs is the similarity metric, defined as:

\begin{equation}
    \begin{split}
     sim = \sum_{i=1}^{k} ({\lambda}_{1i} - {\lambda}_{2i})^2, with 
     \min\bigl\{ \frac{\sum_{i=1}^{k} ({\lambda}_{1i})}{\sum_{i=1}^{k} ({\lambda}_{2i})} > 0.9\bigl\}
    \end{split}
\end{equation}

where $sim\in[0, +\infty)$ for the two sub-graphs $\mathbf{SG}_{1}$ and $\mathbf{SG}_{2}$ and  

\begin{equation}
     \lim_{sim\to0} f({SG}_{1},{SG}_{2}) 
\end{equation}

suggests that the two sub-graphs are similar. The model transferability is determined by considering the improvement in performance (in terms of precision and recall) between the sub-graph on which the model was trained and any other sub-graph it was tested upon:

\begin{equation}
    \mathbf{SG}^{pre}_{i} / \mathbf{SG}^{pre}_{j}
\end{equation}

To this end, we compute the Spearman's rank correlation coefficients ($r_s$) between pairs of similarity and precision or recall ratios. Given the non-parametric nature of our data, we compute the Spearman's rank correlation coefficients ($r_s$) between the scores (Bonferroni-Holm~\cite{Benjamini1995,Benjamini2000} corrected, to guard against over-testing the data). With respect to precision, we observe a significant relationship between the sub-graph similarity $sim$ and relative performance difference for $\mathbf{SG}_{1}$ and all other sub-graph models, $r = -.75, p < .05$. All other tests were not statistically significant. When we consider recall, sub-graph similarity $sim$ was significantly correlated with relative performance difference for several sub-graphs, specifically for $\mathbf{SG}_{1}$ \statreport{r = -.92, p < .01}, $\mathbf{SG}_{2}$ \statreport{r = -.75, p < .05} and $\mathbf{SG}_{4}$ \statreport{r = -.78, p < .01}. This preliminary evidence suggests a connection between the topological similarity of sub-graphs and the ability of the models to generalise to other sub-graphs. 

\subsubsection{Hierarchical Model}
\label{sssec:hierarchical_model}
Next, we consider an ensemble learning extension to our original architecture. Once all the instances of the sub-graph models SG$_{1-k}$ have been fine-tuned and have plateaued in terms of performance, we combine them under a hierarchical structure. To this end, we aggregate the predictions produced by all models and validate them against one sub-graph at a time. This is possible because each of the models is trained on a different set of hyper-parameters and separate graph data, thus allowing them to predict results based on different patterns. The Hierarchical Model (HM) uses as input the predicted labels from each of the $k$ trained models and it compares them against the original target labels. The architecture of the HM consists of three consecutive fully connected linear layers with intermediate ReLU activation functions and a Sigmoid after the last layer (Figure~\ref{fig:model-architecture}). Considering the small size of the sub-graphs and their binary inputs, this is a very efficient and low-energy profile solution. As we demonstrate in Sec.~\ref{sssec:hierarchical_model_results}, their combined use achieves significant performance improvements w.r.t the target metrics.

\begin{figure}[!t]
    \centering
    \includegraphics[width=0.98\linewidth]{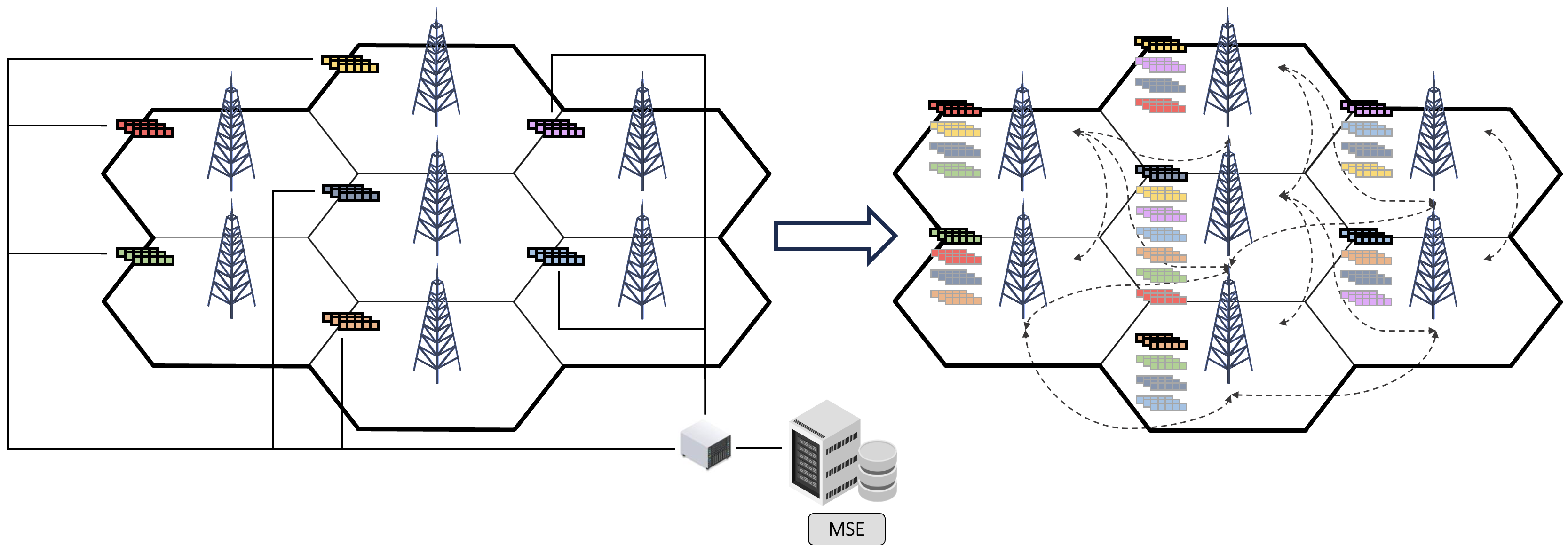}
    \caption{System design flow of LightningNet. Each antenna in the cellular network provides the hourly collected KPIs to the cloud server. Through message passing, neighbouring antennas are sharing information and the final embeddings are calculated through an aggregation function from the features of all neighbours.} 
    \vspace{-0.3cm}
    \label{fig:system-overview}
\end{figure}

\subsubsection{System Design Overview}
\label{sec:system}

Figure~\ref{fig:system-overview} shows an example of how the proposed system design could be deployed in a cellular network, where the antenna cells are organized in a hexagonal deployment. In this scenario, the hourly KPIs are transferred to a local hub and then forwarded to the cloud server. Then, LightningNet provides accurate forecasts about the future cell workloads, where each of the antenna cells collates information as to the status linked to a number of neighbouring cells. The message-passing through these cells is limited to depth $d = 1$, since the introduction of the graph partition with small-sized sub-graphs allows us to traverse through the entire input graph with just one GCN Layer per sub-classifier. Subsequently, each antenna's main computing device, assisted by edge-processing hardware hosting the LightningNet model, receives the incoming antenna KPIs in a streamline fashion and computes the forecasting coefficients. Once all the sub-classifiers have produced their labels, these are forwarded to the cloud server where the HM combines the forecasts for all sub-graph regions to deliver the final predictions for every antenna cell. Each sub-graph is characterized by its own unique training parameters, which results in different training times for the sub-classifiers. Our experimental findings indicate that the average training time of a sub-graph model is approximately an order of a magnitude greater than the average training time of the HM ($\sim$2 min), meaning that the system training time is defined by the slowest sub-classifier. However, the property of the sub-graph models to generalise well across sectors (see Sec.~\ref{ssec:forecasting_results}) reduces the (re)training cycles, thus making the time-delay impact negligible.

\subsubsection{Evaluation Metrics}
\label{sssec:evaluation_measures}

For our model evaluation, we consider the probability of a cell becoming a hot spot in a target prediction horizon and use a decision threshold to separate positive and negative cases. To this end, we consider classical 
information retrieval measures. In particular, given the real-life nature of the task, and the high \textit{operational costs} that false positive cases can incur to MNOs, we define as our primary objective to maximize precision, while maintaining recall at \textit{acceptable} (according to industry and vendor experts) levels. This decision is informed by industrial requirements, which emphasize the need for accurate predictions over costly false positive cases. Moreover, considering the highly imbalanced class distribution observed in our data, we exclude from our analysis performance metrics like accuracy, due to its sensitivity to imbalanced data~\cite{5128907}.

    \section{Experiments}
\label{sec:experiments}


\subsection{Forecasting Results}
\label{ssec:forecasting_results}

\begin{figure*}[!t]
    \captionsetup[subfloat]
    {}
    \centering
    \subfloat[\texttt{hz}=12 hrs]{
    \label{figure:fig1}
    \includegraphics[clip=true, trim=0 0 0 0, width=0.164\textwidth]{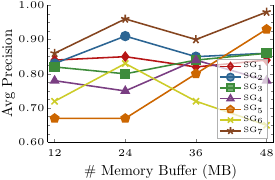}
    }
    \subfloat[\texttt{hz}=24 hrs]{%
    \label{figure:fig2}
    \includegraphics[clip=true, trim=9 0 0 0, width=0.154\textwidth]{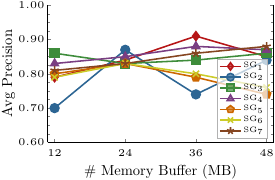}
    }
    \subfloat[\texttt{hz}=48 hrs]{
    \label{figure:fig3}
    \includegraphics[clip=true, trim=9 0 0 0, width=0.154\textwidth]{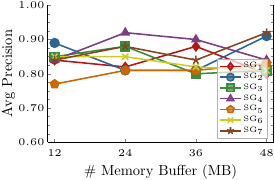}
    }
    \subfloat[\texttt{hz}=12 hrs]{%
    \label{figure:fig4}
    \includegraphics[clip=true, trim=0 0 0 0, width=0.164\textwidth]{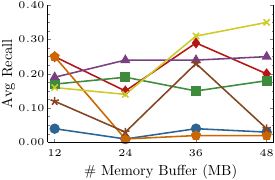}
    }
    \subfloat[\texttt{hz}=24 hrs]{
    \label{figure:fig3a}
    \includegraphics[clip=true, trim=9 0 0 0, width=0.154\textwidth]{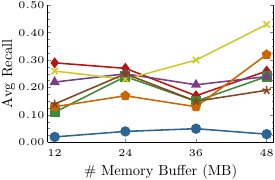}
    }
    \subfloat[\texttt{hz}=48 hrs]{%
    \label{figure:fig4a}
    \includegraphics[clip=true, trim=9 0 0 0, width=0.154\textwidth]{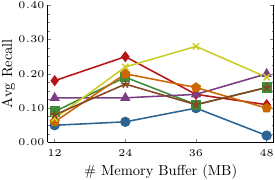}
    }
    \vspace{-0.95em}
    \caption{Average Precision (a,b,c) and Recall (d,e,f) for SG$_{1-7}$, forecasting horizon \texttt{hz} $\in \{12,24,48\}$ and memory buffer \texttt{mb} $\in \{12,24,48\}$ hrs.}
    \label{figure:lineplot-1}
\end{figure*}

\begin{table*}[ht!]
\caption{Performance of LightningNet sub-classifiers(SC$_{1}$-SC$_{7}$) across all sub-graphs (SG$_{1}$-SG$_{7}$), for forecasting horizon \texttt{hz} = 12 hrs and memory buffer \texttt{mb} $\in \{12,24,36,48\}$ hrs. Boldface denotes highest score. Underline indicates performance on the same sub-graph. Average Prec and Rec were calculated using the aggregated confusion matrices.}
\vspace{-1.5em}
\label{table-1}
\vskip 0.1in
\begin{center}
\begin{normalsize}
\begin{sc}
\begin{tabular}{ccp{0.55cm}p{0.55cm}p{0.55cm}p{0.55cm}p{0.55cm}p{0.55cm}p{0.55cm}p{0.55cm}p{0.55cm}p{0.55cm}p{0.55cm}p{0.55cm}p{0.55cm}p{0.8cm}p{1.2cm}p{1.2cm}}
\toprule
& & \multicolumn{2}{c}{SG$_{1}$} & \multicolumn{2}{c}{SG$_{2}$} & \multicolumn{2}{c}{SG$_{3}$} & \multicolumn{2}{c}{SG$_{4}$} & \multicolumn{2}{c}{SG$_{5}$} & \multicolumn{2}{c}{SG$_{6}$} & \multicolumn{2}{c}{SG$_{7}$} & SG$_{1-7}$ & SG$_{1-7}$\cr
\cmidrule(lr){3-4} \cmidrule(lr){5-6} \cmidrule(lr){7-8} \cmidrule(lr){9-10} \cmidrule(lr){11-12} \cmidrule(lr){13-14} \cmidrule(lr){15-16} \cmidrule(lr){17-17} \cmidrule(lr){18-18}
& & Pre & Rec & Pre & Rec & Pre & Rec & Pre & Rec & Pre & Rec & Pre & Rec & Pre & Rec & Avg Pre & Avg Rec \\
\cmidrule(lr){2-18}
\multirow{9}{*}{\rotatebox{90}{\texttt{mb} = 12 hours}} & SC$_{1}$ & \underline{0.82} & \underline{\textbf{0.31}} & 0.76 & 0.15 & 0.86 & 0.27 & 0.83 & 0.23 & \textbf{0.79} & 0.08 & 0.88 & \textbf{0.44} & 0.84 & 0.30 & 0.84 & \textbf{0.25}\\ 
& SC$_{2}$ & \textbf{0.94} & 0.03 & \underline{\textbf{0.82}} & \underline{0.08} & \textbf{0.88} & 0.02 & 0.83 & 0.04 & 0.78 & 0.02 & \textbf{0.96} & 0.05 & 0.74 & 0.08 & 0.83 & 0.04\\ 
& SC$_{3}$ & 0.89 & 0.18 & 0.76 & 0.05 & \underline{0.80} & \underline{0.19} & 0.85 & 0.24 & 0.54 & 0.05 & 0.83 & 0.25 & 0.82 & 0.19 & 0.82 & 0.17\\ 
& SC$_{4}$ & 0.84 & 0.15 & 0.73 & \textbf{0.17} & 0.65 & 0.13 & \underline{0.83} & \underline{0.09} & 0.55 & 0.07 & 0.92 & 0.36 & 0.60 & 0.11 & 0.78 & 0.19\\ 
& SC$_{5}$ & 0.76 & 0.18 & 0.45 & 0.08 & 0.76 & 0.17 & 0.70 & 0.21 & \underline{0.58} & \underline{0.11} & 0.61 & 0.08 & 0.71 & 0.19 & 0.67 & \textbf{0.25}\\ 
& SC$_{6}$ & 0.62 & 0.17 & 0.48 & 0.06 & 0.52 & 0.05 & 0.78 & 0.19 & 0.43 & 0.06 & \underline{0.94} & \underline{0.36} & 0.75 & 0.32 & 0.72 & 0.16\\ 
& SC$_{7}$ & 0.87 & 0.09 & 0.81 & 0.04 & 0.84 & 0.12 & \textbf{0.86} & 0.15 & 0.50 & 0.02 & 0.87 & 0.23 & \underline{\textbf{0.91}} & \underline{0.16} & \textbf{0.86} & 0.12\\
& GCN & 0.37 & 0.10 & 0.00 & 0.00 & 0.42 & 0.12 & 0.26 & 0.02 & 0.00 & 0.00 & 0.00 & 0.00 & 0.19 & 0.08 &  & \\
& LSTM & 0.58 & 0.11 & 0.57 & 0.14 & 0.50 & \textbf{0.39} & 0.68 & \textbf{0.46} & 0.40 & \textbf{0.39} & 0.72 & 0.20 & 0.50 & \textbf{0.46} &  & \\
\midrule
\multirow{9}{*}{\rotatebox{90}{\texttt{mb} = 24 hours}} & SC$_{1}$ & \underline{0.86} & \underline{0.25} & 0.60 & 0.05 & 0.81 & 0.16 & 0.92 & 0.15 & 0.68 & 0.04 & 0.86 & 0.20 & 0.96 & 0.19 & 0.85 & 0.15\\ 
& SC$_{2}$ & 0.67 & 0.00 & \underline{\textbf{1.00}} & \underline{0.04} & 0.83 & 0.03 & 0.83 & 0.01 & \textbf{1.00} & 0.01 & \textbf{1.00} & 0.02 & 0.00 & 0.00 & \textbf{0.91} & 0.01\\ 
& SC$_{3}$ & 0.83 & 0.23 & 0.50 & 0.06 & \underline{0.77} & \underline{0.22} & 0.86 & 0.21 & 0.61 & 0.09 & 0.83 & 0.25 & 0.91 & 0.25 & 0.80 & 0.19\\ 
& SC$_{4}$ & 0.74 & 0.24 & 0.63 & 0.21 & 0.71 & 0.16 & \underline{0.83} & \underline{0.34} & 0.66 & 0.10 & 0.79 & \textbf{0.43} & 0.69 & 0.16 & 0.75 & \textbf{0.24}\\ 
& SC$_{5}$ & \textbf{1.00} & 0.00 & 0.57 & 0.03 & 0.00 & 0.00 & 0.73 & 0.03 & \underline{0.88} & \underline{0.02} & 0.00 & 0.00 & 0.00 & 0.00 & 0.67 & 0.01\\ 
& SC$_{6}$ & 0.62 & 0.11 & 0.88 & 0.07 & 0.80 & 0.10 & 0.87 & 0.13 & 0.49 & 0.04 & \underline{0.95} & \underline{0.33} & 0.98 & 0.19 & 0.83 & 0.14\\ 
& SC$_{7}$ & \textbf{1.00} & 0.03 & 0.00 & 0.00 & \textbf{1.00} & 0.03 & \textbf{1.00} & 0.05 & 0.00 & 0.00 & 0.83 & 0.04 & \underline{\textbf{1.00}} & \underline{0.07} & 0.96 & 0.03\\
& GCN & 0.37 & 0.10 & 0.00 & 0.00 & 0.42 & 0.12 & 0.26 & 0.02 & 0.00 & 0.00 & 0.00 & 0.00 & 0.19 & 0.08 &  & \\
& LSTM & 0.61 & 0.\textbf{49} & 0.36 & \textbf{0.28} & 0.35 & \textbf{0.38} & 0.63 & \textbf{0.47} & 0.46 & \textbf{0.38} & 0.68 & 0.33 & 0.49 & \textbf{0.56} &  & \\
\midrule
\multirow{9}{*}{\rotatebox{90}{\texttt{mb} = 36 hours}} & SC$_{1}$ & \underline{0.83} & \underline{0.33} & 0.72 & 0.25 & 0.79 & 0.27 & 0.91 & 0.30 & 0.68 & 0.11 & 0.87 & 0.49 & 0.82 & 0.26 & 0.82 & 0.29\\ 
& SC$_{2}$ & 0.81 & 0.03 & \underline{\textbf{1.00}} & \underline{0.04} & \textbf{1.00} & 0.02 & 0.74 & 0.05 & \textbf{1.00} & 0.01 & \textbf{0.94} & 0.07 & 0.73 & 0.03 & 0.85 & 0.04\\ 
& SC$_{3}$ & 0.94 & 0.13 & 0.57 & 0.06 & \underline{0.84} & \underline{0.13} & 0.93 & 0.15 & 0.62 & 0.08 & 0.85 & 0.20 & 0.89 & 0.16 & 0.84 & 0.15\\ 
& SC$_{4}$ & 0.87 & 0.23 & 0.71 & 0.14 & 0.82 & 0.15 & \underline{0.85} & \underline{0.37} & 0.60 & 0.08 & 0.87 & 0.44 & 0.91 & 0.21 & 0.84 & 0.24\\ 
& SC$_{5}$ & \textbf{1.00} & 0.01 & 0.83 & 0.01 & 0.67 & 0.01 & 0.78 & 0.01 & \underline{\textbf{1.00}} & \underline{0.02} & 0.63 & 0.01 & 0.50 & 0.01 & 0.80 & 0.02\\ 
& SC$_{6}$ & 0.69 & 0.44 & 0.58 & 0.24 & 0.68 & 0.20 & 0.77 & 0.33 & 0.64 & 0.11 & \underline{0.85} & \underline{\textbf{0.46}} & 0.70 & 0.33 & 0.72 & \textbf{0.31}\\ 
& SC$_{7}$ & 0.86 & 0.17 & 0.87 & 0.07 & 0.83 & 0.11 & \textbf{0.99} & 0.19 & 0.75 & 0.34 & 0.82 & 0.16 & \underline{\textbf{0.99}} & \underline{0.17} & \textbf{0.90} & 0.23\\
& GCN & 0.37 & 0.10 & 0.00 & 0.00 & 0.42 & 0.12 & 0.26 & 0.02 & 0.00 & 0.00 & 0.00 & 0.00 & 0.19 & 0.08 &  & \\
& LSTM & 0.68 & \textbf{0.48} & 0.49 & \textbf{0.40} & 0.50 & \textbf{0.43} & 0.66 & \textbf{0.48} & 0.36 & \textbf{0.39} & 0.00 & 0.00 & 0.53 & \textbf{0.40} &  & \\
\midrule
\multirow{9}{*}{\rotatebox{90}{\texttt{mb} = 48 hours}} & SC$_{1}$ & \underline{0.82} & \underline{0.27} & 0.71 & 0.12 & 0.83 & 0.19 & 0.94 & 0.22 & 0.71 & 0.09 & 0.84 & 0.29 & 0.90 & 0.19 & 0.84 & 0.20\\ 
& SC$_{2}$ & \textbf{1.00} & 0.01 & \underline{0.89} & \underline{0.04} & 0.50 & 0.01 & \textbf{1.00} & 0.01 & 0.40 & 0.00 & 0.95 & 0.09 & 0.70 & 0.02 & 0.86 & 0.03\\ 
& SC$_{3}$ & 0.89 & 0.22 & 0.65 & 0.06 & \underline{0.85} & \underline{0.17} & 0.93 & 0.21 & 0.74 & 0.08 & 0.87 & 0.29 & 0.98 & 0.23 & 0.86 & 0.18\\ 
& SC$_{4}$ & 0.63 & 0.25 & 0.67 & 0.17 & 0.70 & 0.16 & \underline{0.88} & \underline{\textbf{0.35}} & 0.80 & 0.09 & 0.88 & 0.45 & 0.80 & 0.21 & 0.78 & 0.25\\ 
& SC$_{5}$ & 0.82 & 0.02 & \textbf{1.00} & 0.15 & 0.90 & 0.03 & 0.91 & 0.03 & \underline{\textbf{1.00}} & \underline{0.02} & \textbf{1.00} & 0.01 & \textbf{1.00} & 0.02 & \textbf{0.93} & 0.02\\ 
& SC$_{6}$ & 0.72 & 0.30 & 0.51 & \textbf{0.29} & 0.65 & 0.26 & 0.64 & 0.41 & 0.53 & 0.19 & \underline{0.86} & \underline{\textbf{0.55}} & 0.57 & \textbf{0.40} & 0.65 & 0.35\\ 
& SC$_{7}$ & \textbf{1.00} & 0.00 & 0.00 & 0.00 & \textbf{1.00} & 0.03 & \textbf{1.00} & 0.04 & 0.00 & 0.00 & 0.86 & 0.01 & \underline{\textbf{1.00}} & \underline{0.05} & 0.98 & 0.04\\
& GCN & 0.37 & 0.10 & 0.00 & 0.00 & 0.42 & 0.12 & 0.26 & 0.02 & 0.00 & 0.00 & 0.00 & 0.00 & 0.19 & 0.08 &  & \\
& LSTM & 0.64 & \textbf{0.44} & 0.66 & 0.26 & 0.60 & \textbf{0.40} & 0.70 & 0.20 & 0.48 & \textbf{0.30} & 0.55 & 0.26 & 0.70 & \textbf{0.40} &  & \\
\bottomrule
\end{tabular}
\end{sc}
\end{normalsize}
\end{center}
\vskip -0.1in
\end{table*}

\begin{table}[th!]
\caption{Performance of Hierarchical Model for SG$_{1-7}$, for forecasting horizon \texttt{hz}=12 hrs and memory buffer \texttt{mb} $\in \{12,24,36,48\}$ hrs based on aggregated results across.}
\vspace{-1.5em}
\label{table-4}
\vskip 0.1in
\begin{center}
\begin{normalsize}
\begin{sc}
\begin{tabular}{ccccccccc}
\toprule
& \multicolumn{2}{c}{\texttt{mb} = 12h} & \multicolumn{2}{c}{\texttt{mb} = 24h} & \multicolumn{2}{c}{\texttt{mb} = 36h} & \multicolumn{2}{c}{\texttt{mb} = 48h} \cr
\cmidrule(lr){2-3} \cmidrule(lr){4-5} \cmidrule(lr){6-7} \cmidrule(lr){8-9}
& Pre & Rec & Pre & Rec & Pre & Rec & Pre & Rec \\
\midrule
SG$_{1}$ & 1.00 & 0.13 & 1.00 & 0.12 & 0.95 & 0.23 & 0.90 & 0.21\\
SG$_{2}$ & 0.49 & 0.09 & 0.63 & 0.08 & 0.70 & 0.10 & 1.00 & 0.04\\
SG$_{3}$ & 1.00 & 0.16 & 1.00 & 0.05 & 1.00 & 0.11 & 0.88 & 0.11\\
SG$_{4}$ & 1.00 & 0.10 & 1.00 & 0.02 & 1.00 & 0.21 & 1.00 & 0.15\\
SG$_{5}$ & 0.00 & 0.00 & 0.00 & 0.00 & 0.00 & 0.00 & 0.00 & 0.00\\
SG$_{6}$ & 1.00 & 0.02 & 1.00 & 0.07 & 0.89 & 0.28 & 1.00 & 0.26\\
SG$_{7}$ & 1.00 & 0.13 & 1.00 & 0.10 & 1.00 & 0.16 & 1.00 & 0.13\\
\bottomrule
\end{tabular}
\end{sc}
\end{normalsize}
\end{center}
\vskip -0.1in
\end{table}

\subsubsection{Preliminaries}
\label{sssec:results_preliminaries}
In what follows, we evaluate the performance of our model against two state-of-the-art baselines. To this end, we carry out a series of experiments and consider different forecasting horizon and memory buffer size configurations for the RNN component of our architecture. The goal is to demonstrate that the proposed model has a robust performance under different conditions, regardless of the input graph or the observed temporal dynamics. 
Last, we follow a \mbox{80\%-10\%-10\%} split for training, validation and testing in all experiments (see Sec.~\ref{sssec:baselines}).

In our setup, each $k_{th}$ sub-classifier is trained on a different sub-graph (SG$_{1-k}$) after applying an optimization step. The fine-tuning is performed separately for each sub-classifier due to the uniqueness of the input (sub-graph) data, using a held-out validation set, and to determine the optimal hyper-parameters (e.g. number of neurons for the GCN Layer, number of epochs, learning rate). The sub-classifiers then undergo a cross-evaluation process, where they tested against the sub-graph they were trained, as well as every other sub-graph in our dataset. The importance of transferability comes in play in the next step, the hierarchical model. The hierarchical model uses as input the output labels of all the sub-classifiers and tests them against the original labels. This is an attempt to further increase the target performance metric i.e. the model's precision (see Sec.~\ref{sssec:evaluation_measures}).

\subsubsection{LightningNet}
\label{sssec:lightningnet_results}

The performance of the sub-classifiers trained on SG$_{1-7}$ provides an insight to the spatio-temporal patterns in our dataset. First, we see in Figures ~\ref{figure:fig4}-\ref{figure:fig4a} that there is a high variance in recall from sub-graph to sub-graph. This can be attributed to the level of difference in the spatio-temporal dynamics between the sub-graphs, in combination with the total number of positive cases that each of them has. Moreover, we note that the memory buffer with a size of 12 time steps yields the worst and more inconsistent results, which indicates the necessity for larger memory buffer.

When comparing the performance of all models in Table~\ref{table-1} (\texttt{hz}=12), we note that LightningNet outperforms both LSTM and GNN baselines. More specifically, the GNN results are characterised by high variance and poor performance, with precision dropping bellow 50\%. With respect to the LSTM baseline, the results appear to be slightly better compared to the GNN, reaching high precision with relatively high recall, in most cases. Nevertheless, LightningNet appears to outperform both LSTM and GNN models in our target metric (i.e. precision), across all runs. Additionally, the results for other forecasting horizons (i.e. 24, 48) appear to be comparable and are characterised by less variance (as shown by Fig.~\ref{figure:lineplot-1}), with a small degradation observed in recall; therefore, we omit them due to space limitations. 

\begin{figure}[!h]
    \captionsetup[subfloat]
    {}
    \centering
    \subfloat[Precision]{
    \label{figure:barplot-20}
    \includegraphics[clip=true, trim=6 0 0 0, width=0.48\linewidth]{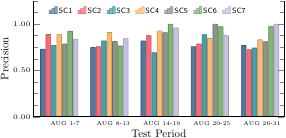}
    }
    \subfloat[Recall]{%
    \label{figure:barplot-21}
    \includegraphics[clip=true, trim=6 0 0 0, width=0.48\linewidth]{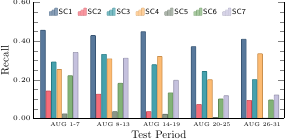}
    }
    \vspace{-0.95em}
    \caption{Sub-Classifier performance metrics across different test periods.}
    \label{figure:barplot-20}
\end{figure}

\subsubsection{Hierarchical Model}
\label{sssec:hierarchical_model_results}
As discussed in Section~\ref{sssec:model_architecture}, the proposed hierachical architecture is similar to the ensemble learning approach, where instead of relying on one ML algorithm to make decisions, we combine the output of several sub-classifiers by learning an ensemble of models and using them in combination. To evaluate the HC, we use the results from the prediction horizon \texttt{hz}=12. The goal here is to make a trade-off between precision and recall, but with an emphasis on maximizing precision (i.e. our target metric).

In Table~\ref{table-4}, we see that for the majority of experiments the HC achieves a higher performance in terms of precision, with a low to average loss in recall compared to the sub-classifiers. The overall performance of the HC is heavily dependent on the findings of the sub-classifiers. Comparing the results of Tables ~\ref{table-1} and ~\ref{table-4}, we can see that training the HC on ${SG}_{5}$ is unfeasible due to sub-optimal performances of all SCs on that particular sub-graph. A similar phenomenon can be observed, to a lesser extent, in the case of ${SG}_{2}$. Another requirement for the smooth operation of the HC, is the level of correlation on the prediction of active cases by the SCs, since this module has a similar behavior to a voting system. In all cases where the HC has provided poor results compared to the corresponding SC, our pipeline will select the findings of the latter as its final output.

\subsubsection{Network Generalization}
\label{sssec:network_generalization}
Here, we consider the proposed model's generalization, an essential consideration  to ensure adaptability and prevent over-fitting. To this end, we use a month's data to train the sub-classifiers, while for inference we evaluate the model's performance on different test sets drawn from the next month's KPIs. We note that the train and test sets are disjoint and we maintain the chronological order of the data. Figure ~\ref{figure:barplot-20} shows a consistent trend - the model maintains optimal performance throughout the dataset, with very small fluctuations. Notably, the model achieves such performance without need of further training. This resilience to temporal dynamics demonstrates that our model indeed is capable of generalizing. The resulting lack of necessity for frequent retraining gives it an advantage, especially when using low-power edge devices, where resources are limited.

\subsection{Performance Analysis on Edge Devices}
\label{ssec:jetson_performance}
An advantage of the proposed architecture is that it addresses the challenge of predicting hot spots in a decentralized manner, while allowing the integration of hardware accelerators (MPSoC/FPGA) with edge devices. This is achieved by deploying lightweight model summaries of graph-based algorithms at the edge. These summaries reduce data size processed per edge device, thereby refining local models and adapting for real-time, edge-based decision-making. Additionally, only local model and data summaries are transmitted, optimizing bandwidth and edge resources utilization, while ensuring privacy. 

For our performance analysis, we chose a widely adopted System on Chip platform, the Jetson Nano Xavier with the Ubuntu 20.04 distribution. This chip contains 6 low power ARM64 CPUs and a cuda based GPU, using 6 GBs of shared memory, and a maximum power consumption of only 10 Watts. An additional benefit of using the Jetson as an edge device is its ability to support both training and inference tasks. Finally, all cloud simulations were carried out on a server running Ubuntu 18.04.2 LTS, with 2 $\times$ Intel(R) Xeon(R) CPU E5-2660 v4 at 2.0 GHz, 
512Gb RAM and 7 $\times$ GeForce GTX 2080 TI (11 Gb).

\subsubsection{Edge Devices versus Cloud}
\label{sssec:Jetson_vs_server performance}
To contrast the hardware performance of the Jetson with the cloud server, we repeat the analysis of the model's computational overhead and compare resource usage for different memory buffer sizes and number of neurons in the model architecture. We are interested in observing how efficient the edge device is compared to the cloud server and what is the difference in energy cost, when considering the proposed model architecture. Memory-wise, the Jetson benefits from shared memory for CPU and GPU operations. As a result, the data to be processed do not have to be copied to the VRAM, therefore allocating less total RAM. 

\begin{figure}[!t]
    \captionsetup[subfloat]
    {}
    \centering
    \subfloat[]{
    \label{figure:barplot-10}
    \includegraphics[clip=true, trim=0 0 0 0, width=0.43\linewidth]{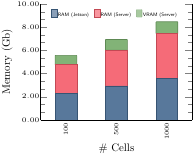}
    }
    \subfloat[]{%
    \label{figure:barplot-13}
    \includegraphics[clip=true, trim=0 0 0 0, width=0.53\linewidth]{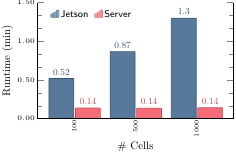}
    }
    \vspace{-0.95em}
    \caption{Performance comparison between Cloud Server and Jetson platform. Figure(a) shows the average memory usage of each platform for varying number of cells. Figure(b) shows the average runtime per epoch.}
    \label{figure:lineplot-3}
\end{figure}

\begin{figure}[!t]
    \captionsetup[subfloat]
    {}
    \centering
    \subfloat[]{
    \label{figure:barplot-17}
    \includegraphics[clip=true, trim=0 0 0 0, width=0.45\linewidth]{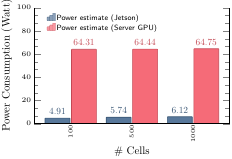}
    }
    \subfloat[]{%
    \label{figure:barplot-19}
    \includegraphics[clip=true, trim=0 0 0 0, width=0.51\linewidth]{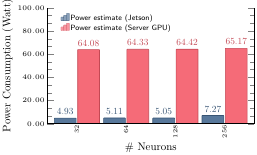}
    }
    \vspace{-0.95em}
    \caption{Average Power Estimate consumption of  Server GPU unit (red) and Jetson platform(blue).}
    \label{figure:lineplot-2}
\end{figure}

As shown in Figure~\ref{figure:barplot-10}, the total memory of the Jetson's RAM module (blue stack) is using less memory than the cloud RAM module (red and green stacks), while requiring additional memory for the GPU device. Due to RAM limitations, only graphs with up to 1000 cells can be processed on the edge device. Comparing runtime (Figure~\ref{figure:barplot-13}), the Jetson platform falls behind due to its less powerful hardware profile, being several times slower in the 1000 cells experiments (the difference in performance increases linearly with the graph size). However, we note that this difference in computational time is in the order of minutes. Additionally, the cloud server retrieves the aggregated data feeds from the mobile network with a latency of several hours, something which will not occur if the chip running on the edge can tap directly to the KPI feeds. Last, we report our results on the Jetson vs. cloud server GPU power consumption (Figures~\ref{figure:barplot-17} and~\ref{figure:barplot-19}). More specifically, the GPU alone consumes 64-150 Watts during training or inference time. On the contrary, the Jetson platform can operate by consuming significantly less power, with a total of 4.6-8.8 Watt across all experiments, which is only a fraction of the power consumed by the cloud server.

\section{Computational Requirements} 
\label{ssec:computational_overhead}
As a side contribution, we look into the computational demands of the proposed solution, an important factor in determining the optimal configuration settings of our solution. More specifically, any parameter fine-tuning (e.g., number of nodes) may impact the computational time, memory requirements, resource drain, power consumption, or even forecasting performance. To this end, we examine the behaviour of the sub-classifier with varying number of neurons, nodes and memory buffer sizes. Considering that the central module of our edge-ML pipeline is the sub-graph, analysing its computational requirements will indicate potential limitations requiring consideration, ultimately informing the final architecture.

\subsection{Training Performance vs. Graph Size}
\label{sssec:cell_config} 
\begin{figure}[!t]
    \captionsetup[subfloat]
    {}
    \centering
    \subfloat[Memory Usage (GB)]{
    \label{figure:barplot-1}
    \includegraphics[clip=true, trim=6 0 0 0, width=0.48\linewidth]{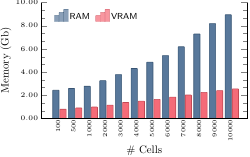}
    }
    \subfloat[Runtime (Minutes)]{
    \label{figure:barplot-2}
    \includegraphics[clip=true, trim=6 0 0 0, width=0.48\linewidth]{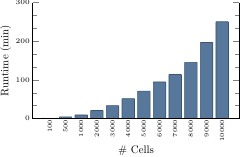}
    }
    \vspace{-0.95em}
    \caption{Cloud Server performance on experiments with varying number of cells showing average memory usage (a) and average runtime per epoch (b).}
    \label{figure:cloud-server}
\end{figure}

We, first, analyse the model's performance at training time, as a function of the graph size. We use a configuration with 256 neurons for the GCN part and a memory buffer of 48 hours for the RNN part, and experiment with a node count that ranges between $100-10,000$, with a step of 1000. This configuration of neurons and memory buffer ensures a good performance for our model in greater node counts. Our trials indicate a linear increase for both RAM and VRAM requirements, as indicated by Figure ~\ref{figure:barplot-1}, and cubic-like increase in execution time, as shown in Figure ~\ref{figure:barplot-2}. These findings suggest that the execution time can become prohibitively high when using graphs with a very large number of nodes/cells.   

\subsection{Training Performance for Different Model Configurations}
\label{sssec:model_config} 
Next, we examine how the memory buffer impacts the model training performance. In our analysis, we consider different memory buffer sizes, ranging from 12 to 48, while keeping constant the graph size and neurons count. Increasing the memory buffer size results in slight variations in RAM consumption, process run-time, and a moderate increase in VRAM usage. Additionally, we replicate our experiments by altering the neuron count from 32 to 256. When increasing the neurons in our model, we observe small increments in RAM consumption and run-time requirements. Furthermore, VRAM usage increases linearly to the neurons.

\subsection{Computational Overhead}
\label{sssec:criticism}
Increasing any of the aforementioned parameters directly translates in an increase in both hardware and training runtime requirements for our model. As previously discussed in Section ~\ref{sssec:model_architecture}, when operating on a bigger graph, we need to fit more complex relations, thus increasing our GCN architectural requirements. For certain datasets, the proposed model may even perform better when equipped with a larger memory buffer. As a result, the total increase in requirements will follow polynomial complexity, making it prohibitive to train and execute on graphs of extreme sizes. On the contrary, the proposed use of lightweight sub-graph models can be faster and less resource-hungry, while also producing accurate forecasts. 
\section{Conclusions}
\label{sec:conclusion}
In this work, we presented LightningNet, the first of its kind distributed ML framework for cellular network performance forecasting, that can capture both spatial and temporal patterns. We demonstrated that our model, with its lightweight architecture, can address the main challenges introduced by large-scale data streams, both in terms of training/execution runtimes and computational resource requirements, thanks to the incorporation of the graph split. Given that the importance of our method lies in the accurate and timely forecasting of hot spots, as well as avoiding costly interventions due to false positive cases, we set as our primary metric precision, while maintaining recall at acceptable levels. Our experiments indicated that LightningNet's performance surpasses that of the baselines and, in the majority of cases, maximizes precision.

Additionally, we showed that LightningNet generalises quite well across sectors, with high transferability, meaning its performance is stable when applied to different sub-graphs than the ones that it was trained on. This can lead to a reduction in the training cycles and energy cost for productising such a forecasting solution. Furthermore, it creates an opportunity for introducing more advanced, ensemble learning solutions. With the implementation of the hierarchical module, we achieved a small trade-off between precision and recall, in the attempt to maximize the former. An added benefit of the proposed architecture is that it can be deployed at the edge using low-power profile devices like MPSoCs, making efficient use of the available bandwidth, and supporting privacy. In conclusion, this work introduces an lightweight solution to cellular network performance forecasting, offering efficiency, accuracy, and paving the way for effective edge-based AI deployment.

\bibliographystyle{ACM-Reference-Format}
\bibliography{main}


\begin{thebibliography}{63}


\ifx \showCODEN    \undefined \def \showCODEN     #1{\unskip}     \fi
\ifx \showDOI      \undefined \def \showDOI       #1{#1}\fi
\ifx \showISBNx    \undefined \def \showISBNx     #1{\unskip}     \fi
\ifx \showISBNxiii \undefined \def \showISBNxiii  #1{\unskip}     \fi
\ifx \showISSN     \undefined \def \showISSN      #1{\unskip}     \fi
\ifx \showLCCN     \undefined \def \showLCCN      #1{\unskip}     \fi
\ifx \shownote     \undefined \def \shownote      #1{#1}          \fi
\ifx \showarticletitle \undefined \def \showarticletitle #1{#1}   \fi
\ifx \showURL      \undefined \def \showURL       {\relax}        \fi
\providecommand\bibfield[2]{#2}
\providecommand\bibinfo[2]{#2}
\providecommand\natexlab[1]{#1}
\providecommand\showeprint[2][]{arXiv:#2}

\bibitem[Agarwal(2008)]%
        {Agarwal2008}
\bibfield{author}{\bibinfo{person}{N. Agarwal}.}
  \bibinfo{year}{2008}\natexlab{}.
\newblock \bibinfo{title}{Wcdma: Kpi analysis \& optimization}.
\newblock \bibinfo{howpublished}{\textit{Nokia Technologies Co., Ltd.}}.
\newblock


\bibitem[Alam(2013)]%
        {Alam2013}
\bibfield{author}{\bibinfo{person}{Md.~Ariful Alam}.}
  \bibinfo{year}{2013}\natexlab{}.
\newblock \showarticletitle{Mobile Network Planning and KPI Improvement
  (Dissertation)}.
\newblock


\bibitem[Atwood and Towsley(2016)]%
        {Atwood2016}
\bibfield{author}{\bibinfo{person}{James Atwood} {and} \bibinfo{person}{Don
  Towsley}.} \bibinfo{year}{2016}\natexlab{}.
\newblock \showarticletitle{Diffusion-Convolutional Neural Networks}. In
  \bibinfo{booktitle}{\emph{Advances in Neural Information Processing
  Systems}}, \bibfield{editor}{\bibinfo{person}{D.~Lee},
  \bibinfo{person}{M.~Sugiyama}, \bibinfo{person}{U.~Luxburg},
  \bibinfo{person}{I.~Guyon}, {and} \bibinfo{person}{R.~Garnett}} (Eds.),
  Vol.~\bibinfo{volume}{29}. \bibinfo{publisher}{Curran Associates, Inc.}
\newblock
\urldef\tempurl%
\url{https://proceedings.neurips.cc/paper/2016/file/390e982518a50e280d8e2b535462ec1f-Paper.pdf}
\showURL{%
\tempurl}


\bibitem[Battaglia et~al\mbox{.}(2018)]%
        {Battaglia2018}
\bibfield{author}{\bibinfo{person}{Peter~W. Battaglia},
  \bibinfo{person}{Jessica~B. Hamrick}, \bibinfo{person}{Victor Bapst},
  \bibinfo{person}{Alvaro Sanchez{-}Gonzalez},
  \bibinfo{person}{Vin{\'{\i}}cius~Flores Zambaldi}, \bibinfo{person}{Mateusz
  Malinowski}, \bibinfo{person}{Andrea Tacchetti}, \bibinfo{person}{David
  Raposo}, \bibinfo{person}{Adam Santoro}, \bibinfo{person}{Ryan Faulkner},
  \bibinfo{person}{{\c{C}}aglar G{\"{u}}l{\c{c}}ehre},
  \bibinfo{person}{H.~Francis Song}, \bibinfo{person}{Andrew~J. Ballard},
  \bibinfo{person}{Justin Gilmer}, \bibinfo{person}{George~E. Dahl},
  \bibinfo{person}{Ashish Vaswani}, \bibinfo{person}{Kelsey~R. Allen},
  \bibinfo{person}{Charles Nash}, \bibinfo{person}{Victoria Langston},
  \bibinfo{person}{Chris Dyer}, \bibinfo{person}{Nicolas Heess},
  \bibinfo{person}{Daan Wierstra}, \bibinfo{person}{Pushmeet Kohli},
  \bibinfo{person}{Matthew~M. Botvinick}, \bibinfo{person}{Oriol Vinyals},
  \bibinfo{person}{Yujia Li}, {and} \bibinfo{person}{Razvan Pascanu}.}
  \bibinfo{year}{2018}\natexlab{}.
\newblock \showarticletitle{Relational inductive biases, deep learning, and
  graph networks}.
\newblock \bibinfo{journal}{\emph{CoRR}}  \bibinfo{volume}{abs/1806.01261}
  (\bibinfo{year}{2018}).
\newblock
\showeprint[arXiv]{1806.01261}
\urldef\tempurl%
\url{http://arxiv.org/abs/1806.01261}
\showURL{%
\tempurl}


\bibitem[Benjamini and Hochberg(1995)]%
        {Benjamini1995}
\bibfield{author}{\bibinfo{person}{Y. Benjamini} {and} \bibinfo{person}{Y.
  Hochberg}.} \bibinfo{year}{1995}\natexlab{}.
\newblock \showarticletitle{Controlling the false discovery rate - a practical
  and powerful approach to multiple testing}.
\newblock \bibinfo{journal}{\emph{Journal of the Royal Statistical Society,
  Series B}} \bibinfo{volume}{57}, \bibinfo{number}{1} (\bibinfo{year}{1995}),
  \bibinfo{pages}{289--300}.
\newblock


\bibitem[Benjamini and Hochberg(2000)]%
        {Benjamini2000}
\bibfield{author}{\bibinfo{person}{Y. Benjamini} {and} \bibinfo{person}{Y.
  Hochberg}.} \bibinfo{year}{2000}\natexlab{}.
\newblock \showarticletitle{On the adaptive control of the false discovery fate
  in multiple testing with independent statistics}.
\newblock \bibinfo{journal}{\emph{Journal of Educational and Behavioral
  Statistics}} \bibinfo{volume}{25}, \bibinfo{number}{1}
  (\bibinfo{year}{2000}), \bibinfo{pages}{60--83}.
\newblock


\bibitem[Cao et~al\mbox{.}(2018)]%
        {Cao2018}
\bibfield{author}{\bibinfo{person}{Wei Cao}, \bibinfo{person}{Dong Wang},
  \bibinfo{person}{Jian Li}, \bibinfo{person}{Hao Zhou}, \bibinfo{person}{Yitan
  Li}, {and} \bibinfo{person}{Lei Li}.} \bibinfo{year}{2018}\natexlab{}.
\newblock \showarticletitle{BRITS: Bidirectional Recurrent Imputation for Time
  Series}. In \bibinfo{booktitle}{\emph{Proceedings of the 32nd International
  Conference on Neural Information Processing Systems}} (Montr\'{e}al, Canada)
  \emph{(\bibinfo{series}{NIPS'18})}. \bibinfo{publisher}{Curran Associates
  Inc.}, \bibinfo{address}{Red Hook, NY, USA}, \bibinfo{pages}{6776–6786}.
\newblock


\bibitem[Cho et~al\mbox{.}(2014)]%
        {ChoMGBSB14}
\bibfield{author}{\bibinfo{person}{Kyunghyun Cho}, \bibinfo{person}{Bart van
  Merrienboer}, \bibinfo{person}{{\c{C}}aglar G{\"{u}}l{\c{c}}ehre},
  \bibinfo{person}{Fethi Bougares}, \bibinfo{person}{Holger Schwenk}, {and}
  \bibinfo{person}{Yoshua Bengio}.} \bibinfo{year}{2014}\natexlab{}.
\newblock \showarticletitle{Learning Phrase Representations using {RNN}
  Encoder-Decoder for Statistical Machine Translation}.
\newblock \bibinfo{journal}{\emph{CoRR}}  \bibinfo{volume}{abs/1406.1078}
  (\bibinfo{year}{2014}).
\newblock
\showeprint[arXiv]{1406.1078}
\urldef\tempurl%
\url{http://arxiv.org/abs/1406.1078}
\showURL{%
\tempurl}


\bibitem[Chung et~al\mbox{.}(2014)]%
        {ChungGCB14}
\bibfield{author}{\bibinfo{person}{Junyoung Chung},
  \bibinfo{person}{{\c{C}}aglar G{\"{u}}l{\c{c}}ehre},
  \bibinfo{person}{KyungHyun Cho}, {and} \bibinfo{person}{Yoshua Bengio}.}
  \bibinfo{year}{2014}\natexlab{}.
\newblock \showarticletitle{Empirical Evaluation of Gated Recurrent Neural
  Networks on Sequence Modeling}.
\newblock \bibinfo{journal}{\emph{CoRR}}  \bibinfo{volume}{abs/1412.3555}
  (\bibinfo{year}{2014}).
\newblock
\showeprint[arXiv]{1412.3555}
\urldef\tempurl%
\url{http://arxiv.org/abs/1412.3555}
\showURL{%
\tempurl}


\bibitem[Chung et~al\mbox{.}(2015)]%
        {Chung2015}
\bibfield{author}{\bibinfo{person}{Junyoung Chung}, \bibinfo{person}{Kyle
  Kastner}, \bibinfo{person}{Laurent Dinh}, \bibinfo{person}{Kratarth Goel},
  \bibinfo{person}{Aaron Courville}, {and} \bibinfo{person}{Yoshua Bengio}.}
  \bibinfo{year}{2015}\natexlab{}.
\newblock \showarticletitle{A Recurrent Latent Variable Model for Sequential
  Data}. In \bibinfo{booktitle}{\emph{Proceedings of the 28th International
  Conference on Neural Information Processing Systems - Volume 2}} (Montreal,
  Canada) \emph{(\bibinfo{series}{NIPS'15})}. \bibinfo{publisher}{MIT Press},
  \bibinfo{address}{Cambridge, MA, USA}, \bibinfo{pages}{2980–2988}.
\newblock


\bibitem[Cui et~al\mbox{.}(2018)]%
        {https://doi.org/10.48550/arxiv.1802.07007}
\bibfield{author}{\bibinfo{person}{Zhiyong Cui}, \bibinfo{person}{Kristian
  Henrickson}, \bibinfo{person}{Ruimin Ke}, \bibinfo{person}{Ziyuan Pu}, {and}
  \bibinfo{person}{Yinhai Wang}.} \bibinfo{year}{2018}\natexlab{}.
\newblock \bibinfo{title}{Traffic Graph Convolutional Recurrent Neural Network:
  A Deep Learning Framework for Network-Scale Traffic Learning and
  Forecasting}.
\newblock
\newblock
\urldef\tempurl%
\url{https://doi.org/10.48550/ARXIV.1802.07007}
\showDOI{\tempurl}


\bibitem[Dalgkitsis et~al\mbox{.}(2018)]%
        {Dalgkitsis2018}
\bibfield{author}{\bibinfo{person}{Anestis Dalgkitsis},
  \bibinfo{person}{Malamati Louta}, {and} \bibinfo{person}{George~T.
  Karetsos}.} \bibinfo{year}{2018}\natexlab{}.
\newblock \showarticletitle{Traffic Forecasting in Cellular Networks Using the
  LSTM RNN}. In \bibinfo{booktitle}{\emph{Proceedings of the 22nd Pan-Hellenic
  Conference on Informatics}} (Athens, Greece) \emph{(\bibinfo{series}{PCI
  '18})}. \bibinfo{publisher}{Association for Computing Machinery},
  \bibinfo{address}{New York, NY, USA}, \bibinfo{pages}{28–33}.
\newblock
\showISBNx{9781450366106}
\urldef\tempurl%
\url{https://doi.org/10.1145/3291533.3291540}
\showDOI{\tempurl}


\bibitem[Duan et~al\mbox{.}(2016)]%
        {Duan2016}
\bibfield{author}{\bibinfo{person}{Yanjie Duan}, \bibinfo{person}{Yisheng
  L.V.}, {and} \bibinfo{person}{Fei-Yue Wang}.}
  \bibinfo{year}{2016}\natexlab{}.
\newblock \showarticletitle{Travel time prediction with LSTM neural network}.
  In \bibinfo{booktitle}{\emph{2016 IEEE 19th International Conference on
  Intelligent Transportation Systems (ITSC)}}. \bibinfo{pages}{1053--1058}.
\newblock
\urldef\tempurl%
\url{https://doi.org/10.1109/ITSC.2016.7795686}
\showDOI{\tempurl}


\bibitem[Ericsson(2014)]%
        {Ericsson2014}
\bibfield{author}{\bibinfo{person}{Ericsson}.} \bibinfo{year}{2014}\natexlab{}.
\newblock \bibinfo{title}{Measuring and improving network performance}.
\newblock \bibinfo{howpublished}{\textit{White paper}}.
\newblock


\bibitem[Fu et~al\mbox{.}(2016)]%
        {RNN_traffic}
\bibfield{author}{\bibinfo{person}{Rui Fu}, \bibinfo{person}{Zuo Zhang}, {and}
  \bibinfo{person}{Li Li}.} \bibinfo{year}{2016}\natexlab{}.
\newblock \showarticletitle{Using LSTM and GRU neural network methods for
  traffic flow prediction}. In \bibinfo{booktitle}{\emph{2016 31st Youth
  Academic Annual Conference of Chinese Association of Automation (YAC)}}.
  \bibinfo{pages}{324--328}.
\newblock
\urldef\tempurl%
\url{https://doi.org/10.1109/YAC.2016.7804912}
\showDOI{\tempurl}


\bibitem[Hamed et~al\mbox{.}(1995)]%
        {Hamed1995}
\bibfield{author}{\bibinfo{person}{Mohammad~M. Hamed},
  \bibinfo{person}{Hashem~R. Al-Masaeid}, {and} \bibinfo{person}{Z~M~Bani
  Said}.} \bibinfo{year}{1995}\natexlab{}.
\newblock \showarticletitle{Short-Term Prediction of Traffic Volume in Urban
  Arterials}.
\newblock \bibinfo{journal}{\emph{Journal of Transportation Engineering-asce}}
  \bibinfo{volume}{121} (\bibinfo{year}{1995}), \bibinfo{pages}{249--254}.
\newblock


\bibitem[Hamrick et~al\mbox{.}(2018)]%
        {Hamrick2018}
\bibfield{author}{\bibinfo{person}{Jessica~B. Hamrick},
  \bibinfo{person}{Kelsey~R. Allen}, \bibinfo{person}{Victor Bapst},
  \bibinfo{person}{Tina Zhu}, \bibinfo{person}{Kevin~R. McKee},
  \bibinfo{person}{Joshua~B. Tenenbaum}, {and} \bibinfo{person}{Peter~W.
  Battaglia}.} \bibinfo{year}{2018}\natexlab{}.
\newblock \showarticletitle{Relational inductive bias for physical construction
  in humans and machines}.
\newblock \bibinfo{journal}{\emph{CoRR}}  \bibinfo{volume}{abs/1806.01203}
  (\bibinfo{year}{2018}).
\newblock
\showeprint[arXiv]{1806.01203}
\urldef\tempurl%
\url{http://arxiv.org/abs/1806.01203}
\showURL{%
\tempurl}


\bibitem[He and Garcia(2009)]%
        {5128907}
\bibfield{author}{\bibinfo{person}{Haibo He} {and} \bibinfo{person}{Edwardo~A.
  Garcia}.} \bibinfo{year}{2009}\natexlab{}.
\newblock \showarticletitle{Learning from Imbalanced Data}.
\newblock \bibinfo{journal}{\emph{IEEE Transactions on Knowledge and Data
  Engineering}} \bibinfo{volume}{21}, \bibinfo{number}{9}
  (\bibinfo{year}{2009}), \bibinfo{pages}{1263--1284}.
\newblock
\urldef\tempurl%
\url{https://doi.org/10.1109/TKDE.2008.239}
\showDOI{\tempurl}


\bibitem[Henaff et~al\mbox{.}(2015)]%
        {Henaff2015}
\bibfield{author}{\bibinfo{person}{Mikael Henaff}, \bibinfo{person}{Joan
  Bruna}, {and} \bibinfo{person}{Yann LeCun}.} \bibinfo{year}{2015}\natexlab{}.
\newblock \showarticletitle{Deep Convolutional Networks on Graph-Structured
  Data}.
\newblock \bibinfo{journal}{\emph{CoRR}}  \bibinfo{volume}{abs/1506.05163}
  (\bibinfo{year}{2015}).
\newblock
\showeprint[arXiv]{1506.05163}
\urldef\tempurl%
\url{http://arxiv.org/abs/1506.05163}
\showURL{%
\tempurl}


\bibitem[Hochreiter and Schmidhuber(1997)]%
        {Hochreiter1997}
\bibfield{author}{\bibinfo{person}{Sepp Hochreiter} {and}
  \bibinfo{person}{J\"{u}rgen Schmidhuber}.} \bibinfo{year}{1997}\natexlab{}.
\newblock \showarticletitle{Long Short-Term Memory}.
\newblock \bibinfo{journal}{\emph{Neural Comput.}} \bibinfo{volume}{9},
  \bibinfo{number}{8} (\bibinfo{date}{nov} \bibinfo{year}{1997}),
  \bibinfo{pages}{1735–1780}.
\newblock
\showISSN{0899-7667}
\urldef\tempurl%
\url{https://doi.org/10.1162/neco.1997.9.8.1735}
\showDOI{\tempurl}


\bibitem[Holma and Toskala(2007)]%
        {Holma2007}
\bibfield{author}{\bibinfo{person}{Harri Holma} {and} \bibinfo{person}{Antti
  Toskala}.} \bibinfo{year}{2007}\natexlab{}.
\newblock \bibinfo{booktitle}{\emph{WCDMA for UMTS: HSPA Evolution and LTE}}.
\newblock \bibinfo{publisher}{John Wiley \& Sons, Inc.},
  \bibinfo{address}{USA}.
\newblock
\showISBNx{047031933X}


\bibitem[Hopfield(1982)]%
        {Hopfield1982}
\bibfield{author}{\bibinfo{person}{J~J Hopfield}.}
  \bibinfo{year}{1982}\natexlab{}.
\newblock \showarticletitle{Neural networks and physical systems with emergent
  collective computational abilities.}
\newblock \bibinfo{journal}{\emph{Proceedings of the National Academy of
  Sciences}} \bibinfo{volume}{79}, \bibinfo{number}{8} (\bibinfo{year}{1982}),
  \bibinfo{pages}{2554--2558}.
\newblock
\urldef\tempurl%
\url{https://doi.org/10.1073/pnas.79.8.2554}
\showDOI{\tempurl}
\showeprint{https://www.pnas.org/doi/pdf/10.1073/pnas.79.8.2554}


\bibitem[Jin et~al\mbox{.}(2018)]%
        {Jin2018}
\bibfield{author}{\bibinfo{person}{Wenwei Jin}, \bibinfo{person}{Youfang Lin},
  \bibinfo{person}{Zhihao Wu}, {and} \bibinfo{person}{Huaiyu Wan}.}
  \bibinfo{year}{2018}\natexlab{}.
\newblock \showarticletitle{Spatio-Temporal Recurrent Convolutional Networks
  for Citywide Short-Term Crowd Flows Prediction}. In
  \bibinfo{booktitle}{\emph{Proceedings of the 2nd International Conference on
  Compute and Data Analysis}} (DeKalb, IL, USA) \emph{(\bibinfo{series}{ICCDA
  2018})}. \bibinfo{publisher}{Association for Computing Machinery},
  \bibinfo{address}{New York, NY, USA}, \bibinfo{pages}{28–35}.
\newblock
\showISBNx{9781450363594}
\urldef\tempurl%
\url{https://doi.org/10.1145/3193077.3193082}
\showDOI{\tempurl}


\bibitem[Jordan(1997)]%
        {Jordan1997}
\bibfield{author}{\bibinfo{person}{Michael~I. Jordan}.}
  \bibinfo{year}{1997}\natexlab{}.
\newblock \showarticletitle{Serial Order: A Parallel Distributed Processing
  Approach}.
\newblock \bibinfo{journal}{\emph{Advances in psychology}}
  \bibinfo{volume}{121} (\bibinfo{year}{1997}), \bibinfo{pages}{471--495}.
\newblock


\bibitem[Kaiping and Hao(2008)]%
        {Kaiping2008}
\bibfield{author}{\bibinfo{person}{X. Kaiping} {and} \bibinfo{person}{G. Hao}.}
  \bibinfo{year}{2008}\natexlab{}.
\newblock \bibinfo{title}{GSM KPI monitoring and improvement guide}.
\newblock \bibinfo{howpublished}{\textit{Huawei Technologies Co., Ltd.}}.
\newblock


\bibitem[Kalafatas and Peeta(2007)]%
        {Kalafatas2007}
\bibfield{author}{\bibinfo{person}{Georgios Kalafatas} {and}
  \bibinfo{person}{Srinivas Peeta}.} \bibinfo{year}{2007}\natexlab{}.
\newblock \showarticletitle{An Exact Graph Structure for Dynamic Traffic
  Assignment: Formulation, Properties, and Computational Experience}. In
  \bibinfo{booktitle}{\emph{Transportation Research Board 86th Annual
  Meeting}}.
\newblock


\bibitem[Kipf and Welling(2016)]%
        {Kipf2016}
\bibfield{author}{\bibinfo{person}{Thomas~N. Kipf} {and} \bibinfo{person}{Max
  Welling}.} \bibinfo{year}{2016}\natexlab{}.
\newblock \showarticletitle{Semi-Supervised Classification with Graph
  Convolutional Networks}.
\newblock \bibinfo{journal}{\emph{CoRR}}  \bibinfo{volume}{abs/1609.02907}
  (\bibinfo{year}{2016}).
\newblock
\showeprint[arXiv]{1609.02907}
\urldef\tempurl%
\url{http://arxiv.org/abs/1609.02907}
\showURL{%
\tempurl}


\bibitem[Koutra et~al\mbox{.}(2011)]%
        {Koutra2011AlgorithmsFG}
\bibfield{author}{\bibinfo{person}{Danai Koutra}, \bibinfo{person}{Ankur~P.
  Parikh}, \bibinfo{person}{Aaditya Ramdas}, {and} \bibinfo{person}{Jing
  Xiang}.} \bibinfo{year}{2011}\natexlab{}.
\newblock \showarticletitle{Algorithms for Graph Similarity and Subgraph
  Matching}.
\newblock


\bibitem[Kreher and Gaenger(2015)]%
        {Kreher2015}
\bibfield{author}{\bibinfo{person}{Ralf Kreher} {and} \bibinfo{person}{Karsten
  Gaenger}.} \bibinfo{year}{2015}\natexlab{}.
\newblock \bibinfo{booktitle}{\emph{Key Performance Indicators and Measurements
  for LTE Radio Network Optimization}}.
\newblock \bibinfo{pages}{267--336}.
\newblock
\urldef\tempurl%
\url{https://doi.org/10.1002/9781118725092.ch4}
\showDOI{\tempurl}


\bibitem[{L. Huawei Technologies Co.}(2006)]%
        {Huawei2006}
\bibfield{author}{\bibinfo{person}{{L. Huawei Technologies Co.}}}
  \bibinfo{year}{2006}\natexlab{}.
\newblock \bibinfo{booktitle}{\emph{HUAWEI RAN KPI for performance
  management.}}
\newblock \bibinfo{type}{{T}echnical {R}eport}.
\newblock


\bibitem[Lai et~al\mbox{.}(2018)]%
        {Lai2018}
\bibfield{author}{\bibinfo{person}{Guokun Lai}, \bibinfo{person}{Wei-Cheng
  Chang}, \bibinfo{person}{Yiming Yang}, {and} \bibinfo{person}{Hanxiao Liu}.}
  \bibinfo{year}{2018}\natexlab{}.
\newblock \showarticletitle{Modeling Long- and Short-Term Temporal Patterns
  with Deep Neural Networks}. In \bibinfo{booktitle}{\emph{The 41st
  International ACM SIGIR Conference on Research \& Development in Information
  Retrieval}} (Ann Arbor, MI, USA) \emph{(\bibinfo{series}{SIGIR '18})}.
  \bibinfo{publisher}{Association for Computing Machinery},
  \bibinfo{address}{New York, NY, USA}, \bibinfo{pages}{95–104}.
\newblock
\showISBNx{9781450356572}
\urldef\tempurl%
\url{https://doi.org/10.1145/3209978.3210006}
\showDOI{\tempurl}


\bibitem[Laiho et~al\mbox{.}(2001)]%
        {Laiho2001}
\bibfield{author}{\bibinfo{person}{Jaana Laiho}, \bibinfo{person}{Achim
  Wacker}, {and} \bibinfo{person}{T. Novosad}.}
  \bibinfo{year}{2001}\natexlab{}.
\newblock \bibinfo{booktitle}{\emph{Radio Network Planning and Optimisation for
  UMTS}}.
\newblock
\showISBNx{0471486531, 978-0471486534}
\urldef\tempurl%
\url{https://doi.org/10.1002/9780470031407}
\showDOI{\tempurl}


\bibitem[Leontiadis et~al\mbox{.}(2017)]%
        {Leontiadis2017}
\bibfield{author}{\bibinfo{person}{Ilias Leontiadis}, \bibinfo{person}{Joan
  Serrà}, \bibinfo{person}{Alessandro Finamore}, \bibinfo{person}{Giorgos
  Dimopoulos}, {and} \bibinfo{person}{Konstantina Papagiannaki}.}
  \bibinfo{year}{2017}\natexlab{}.
\newblock \showarticletitle{The Good, the Bad, and the KPIs: How to Combine
  Performance Metrics to Better Capture Underperforming Sectors in Mobile
  Networks}. In \bibinfo{booktitle}{\emph{2017 IEEE 33rd International
  Conference on Data Engineering (ICDE)}}. \bibinfo{pages}{297--308}.
\newblock
\urldef\tempurl%
\url{https://doi.org/10.1109/ICDE.2017.89}
\showDOI{\tempurl}


\bibitem[Li et~al\mbox{.}(2020)]%
        {Can2020}
\bibfield{author}{\bibinfo{person}{Can Li}, \bibinfo{person}{Lei Bai},
  \bibinfo{person}{Wei Liu}, \bibinfo{person}{Lina Yao}, {and}
  \bibinfo{person}{S.~Travis Waller}.} \bibinfo{year}{2020}\natexlab{}.
\newblock \showarticletitle{Knowledge Adaption for Demand Prediction Based on
  Multi-Task Memory Neural Network}. In \bibinfo{booktitle}{\emph{Proceedings
  of the 29th ACM International Conference on Information \& Knowledge
  Management}} (Virtual Event, Ireland) \emph{(\bibinfo{series}{CIKM '20})}.
  \bibinfo{publisher}{Association for Computing Machinery},
  \bibinfo{address}{New York, NY, USA}, \bibinfo{pages}{715–724}.
\newblock
\showISBNx{9781450368599}
\urldef\tempurl%
\url{https://doi.org/10.1145/3340531.3411965}
\showDOI{\tempurl}


\bibitem[Li et~al\mbox{.}(2015)]%
        {li2015}
\bibfield{author}{\bibinfo{person}{Li Li}, \bibinfo{person}{Ke Xu},
  \bibinfo{person}{Dan Wang}, \bibinfo{person}{Chunyi Peng},
  \bibinfo{person}{Qingyang Xiao}, {and} \bibinfo{person}{Rashid Mijumbi}.}
  \bibinfo{year}{2015}\natexlab{}.
\newblock \showarticletitle{A measurement study on TCP behaviors in HSPA+
  networks on high-speed rails}. In \bibinfo{booktitle}{\emph{2015 IEEE
  Conference on Computer Communications (INFOCOM)}}.
  \bibinfo{pages}{2731--2739}.
\newblock
\urldef\tempurl%
\url{https://doi.org/10.1109/INFOCOM.2015.7218665}
\showDOI{\tempurl}


\bibitem[Li et~al\mbox{.}(2017)]%
        {Yaguang2017}
\bibfield{author}{\bibinfo{person}{Yaguang Li}, \bibinfo{person}{Rose Yu},
  \bibinfo{person}{Cyrus Shahabi}, {and} \bibinfo{person}{Yan Liu}.}
  \bibinfo{year}{2017}\natexlab{}.
\newblock \showarticletitle{Graph Convolutional Recurrent Neural Network:
  Data-Driven Traffic Forecasting}.
\newblock \bibinfo{journal}{\emph{CoRR}}  \bibinfo{volume}{abs/1707.01926}
  (\bibinfo{year}{2017}).
\newblock
\showeprint[arXiv]{1707.01926}
\urldef\tempurl%
\url{http://arxiv.org/abs/1707.01926}
\showURL{%
\tempurl}


\bibitem[Luo et~al\mbox{.}(2018)]%
        {Luo2018}
\bibfield{author}{\bibinfo{person}{Yonghong Luo}, \bibinfo{person}{Xiangrui
  Cai}, \bibinfo{person}{Ying Zhang}, \bibinfo{person}{Jun Xu}, {and}
  \bibinfo{person}{Xiaojie Yuan}.} \bibinfo{year}{2018}\natexlab{}.
\newblock \showarticletitle{Multivariate Time Series Imputation with Generative
  Adversarial Networks}. In \bibinfo{booktitle}{\emph{Proceedings of the 32nd
  International Conference on Neural Information Processing Systems}}
  (Montr\'{e}al, Canada) \emph{(\bibinfo{series}{NIPS'18})}.
  \bibinfo{publisher}{Curran Associates Inc.}, \bibinfo{address}{Red Hook, NY,
  USA}, \bibinfo{pages}{1603–1614}.
\newblock


\bibitem[Lv et~al\mbox{.}(2015)]%
        {SAEs}
\bibfield{author}{\bibinfo{person}{Yisheng Lv}, \bibinfo{person}{Yanjie Duan},
  \bibinfo{person}{Wenwen Kang}, \bibinfo{person}{Zhengxi Li}, {and}
  \bibinfo{person}{Fei-Yue Wang}.} \bibinfo{year}{2015}\natexlab{}.
\newblock \showarticletitle{Traffic Flow Prediction With Big Data: A Deep
  Learning Approach}.
\newblock \bibinfo{journal}{\emph{IEEE Transactions on Intelligent
  Transportation Systems}} \bibinfo{volume}{16}, \bibinfo{number}{2}
  (\bibinfo{year}{2015}), \bibinfo{pages}{865--873}.
\newblock
\urldef\tempurl%
\url{https://doi.org/10.1109/TITS.2014.2345663}
\showDOI{\tempurl}


\bibitem[Ma et~al\mbox{.}(2017)]%
        {Ma2017LearningTA}
\bibfield{author}{\bibinfo{person}{Xiaolei Ma}, \bibinfo{person}{Zhuang Dai},
  \bibinfo{person}{Zhengbing He}, \bibinfo{person}{Jihui Ma},
  \bibinfo{person}{Yong Wang}, {and} \bibinfo{person}{Yunpeng Wang}.}
  \bibinfo{year}{2017}\natexlab{}.
\newblock \showarticletitle{Learning Traffic as Images: A Deep Convolutional
  Neural Network for Large-Scale Transportation Network Speed Prediction}.
\newblock \bibinfo{journal}{\emph{Sensors (Basel, Switzerland)}}
  \bibinfo{volume}{17} (\bibinfo{year}{2017}).
\newblock


\bibitem[Ma et~al\mbox{.}(2015)]%
        {MA2015187}
\bibfield{author}{\bibinfo{person}{Xiaolei Ma}, \bibinfo{person}{Zhimin Tao},
  \bibinfo{person}{Yinhai Wang}, \bibinfo{person}{Haiyang Yu}, {and}
  \bibinfo{person}{Yunpeng Wang}.} \bibinfo{year}{2015}\natexlab{}.
\newblock \showarticletitle{Long short-term memory neural network for traffic
  speed prediction using remote microwave sensor data}.
\newblock \bibinfo{journal}{\emph{Transportation Research Part C: Emerging
  Technologies}}  \bibinfo{volume}{54} (\bibinfo{year}{2015}),
  \bibinfo{pages}{187--197}.
\newblock
\showISSN{0968-090X}
\urldef\tempurl%
\url{https://doi.org/10.1016/j.trc.2015.03.014}
\showDOI{\tempurl}


\bibitem[Mahimkar et~al\mbox{.}(2013)]%
        {Mahimkar2013}
\bibfield{author}{\bibinfo{person}{Ajay Mahimkar}, \bibinfo{person}{Zihui Ge},
  \bibinfo{person}{Jennifer Yates}, \bibinfo{person}{Chris Hristov},
  \bibinfo{person}{Vincent Cordaro}, \bibinfo{person}{Shane Smith},
  \bibinfo{person}{Jing Xu}, {and} \bibinfo{person}{Mark Stockert}.}
  \bibinfo{year}{2013}\natexlab{}.
\newblock \showarticletitle{Robust Assessment of Changes in Cellular Networks}.
  In \bibinfo{booktitle}{\emph{Proceedings of the Ninth ACM Conference on
  Emerging Networking Experiments and Technologies}} (Santa Barbara,
  California, USA) \emph{(\bibinfo{series}{CoNEXT '13})}.
  \bibinfo{publisher}{Association for Computing Machinery},
  \bibinfo{address}{New York, NY, USA}, \bibinfo{pages}{175–186}.
\newblock
\showISBNx{9781450321013}
\urldef\tempurl%
\url{https://doi.org/10.1145/2535372.2535382}
\showDOI{\tempurl}


\bibitem[Mishra(2004)]%
        {mishra2004}
\bibfield{author}{\bibinfo{person}{R.~Ajay Mishra}.}
  \bibinfo{year}{2004}\natexlab{}.
\newblock \bibinfo{booktitle}{\emph{Fundamentals of Cellular Network Planning
  and Optimisation}}.
\newblock \bibinfo{publisher}{John Wiley \& Sons, Ltd}.
\newblock
\showISBNx{9780470862698}
\urldef\tempurl%
\url{https://doi.org/10.1002/0470862696.ch10}
\showDOI{\tempurl}
\showeprint{https://onlinelibrary.wiley.com/doi/pdf/10.1002/0470862696.ch10}


\bibitem[Nika et~al\mbox{.}(2014)]%
        {Nika2014}
\bibfield{author}{\bibinfo{person}{Ana Nika}, \bibinfo{person}{Asad Ismail},
  \bibinfo{person}{Ben Zhao}, \bibinfo{person}{Sabrina Gaito},
  \bibinfo{person}{Gian Rossi}, {and} \bibinfo{person}{Haitao Zheng}.}
  \bibinfo{year}{2014}\natexlab{}.
\newblock \showarticletitle{Understanding Data Hotspots in Cellular Networks}.
  \bibinfo{publisher}{IEEE}.
\newblock
\urldef\tempurl%
\url{https://doi.org/10.4108/icst.qshine.2014.256291}
\showDOI{\tempurl}


\bibitem[Polson and Sokolov(2017)]%
        {POLSON20171}
\bibfield{author}{\bibinfo{person}{Nicholas~G. Polson} {and}
  \bibinfo{person}{Vadim~O. Sokolov}.} \bibinfo{year}{2017}\natexlab{}.
\newblock \showarticletitle{Deep learning for short-term traffic flow
  prediction}.
\newblock \bibinfo{journal}{\emph{Transportation Research Part C: Emerging
  Technologies}}  \bibinfo{volume}{79} (\bibinfo{year}{2017}),
  \bibinfo{pages}{1--17}.
\newblock
\showISSN{0968-090X}
\urldef\tempurl%
\url{https://doi.org/10.1016/j.trc.2017.02.024}
\showDOI{\tempurl}


\bibitem[Rangapuram et~al\mbox{.}(2018)]%
        {Rangapuram2018}
\bibfield{author}{\bibinfo{person}{Syama~Sundar Rangapuram},
  \bibinfo{person}{Matthias Seeger}, \bibinfo{person}{Jan Gasthaus},
  \bibinfo{person}{Lorenzo Stella}, \bibinfo{person}{Yuyang Wang}, {and}
  \bibinfo{person}{Tim Januschowski}.} \bibinfo{year}{2018}\natexlab{}.
\newblock \showarticletitle{Deep State Space Models for Time Series
  Forecasting}. In \bibinfo{booktitle}{\emph{Proceedings of the 32nd
  International Conference on Neural Information Processing Systems}}
  (Montr\'{e}al, Canada) \emph{(\bibinfo{series}{NIPS'18})}.
  \bibinfo{publisher}{Curran Associates Inc.}, \bibinfo{address}{Red Hook, NY,
  USA}, \bibinfo{pages}{7796–7805}.
\newblock


\bibitem[Rumelhart et~al\mbox{.}(1986)]%
        {Rumelhart1986}
\bibfield{author}{\bibinfo{person}{David~E. Rumelhart},
  \bibinfo{person}{Geoffrey~E. Hinton}, {and} \bibinfo{person}{Ronald~J.
  Williams}.} \bibinfo{year}{1986}\natexlab{}.
\newblock \showarticletitle{Learning representations by back-propagating
  errors}.
\newblock \bibinfo{journal}{\emph{Nature}}  \bibinfo{volume}{323}
  (\bibinfo{year}{1986}), \bibinfo{pages}{533--536}.
\newblock


\bibitem[Salinas et~al\mbox{.}(2020)]%
        {Salinas2020}
\bibfield{author}{\bibinfo{person}{David Salinas}, \bibinfo{person}{Valentin
  Flunkert}, \bibinfo{person}{Jan Gasthaus}, {and} \bibinfo{person}{Tim
  Januschowski}.} \bibinfo{year}{2020}\natexlab{}.
\newblock \showarticletitle{DeepAR: Probabilistic forecasting with
  autoregressive recurrent networks}.
\newblock \bibinfo{journal}{\emph{International Journal of Forecasting}}
  \bibinfo{volume}{36}, \bibinfo{number}{3} (\bibinfo{year}{2020}),
  \bibinfo{pages}{1181--1191}.
\newblock
\showISSN{0169-2070}
\urldef\tempurl%
\url{https://doi.org/10.1016/j.ijforecast.2019.07.001}
\showDOI{\tempurl}


\bibitem[Serr\`{a} et~al\mbox{.}(2017)]%
        {Serra2017}
\bibfield{author}{\bibinfo{person}{Joan Serr\`{a}}, \bibinfo{person}{Ilias
  Leontiadis}, \bibinfo{person}{Alexandros Karatzoglou}, {and}
  \bibinfo{person}{Konstantina Papagiannaki}.} \bibinfo{year}{2017}\natexlab{}.
\newblock \showarticletitle{Hot or Not? Forecasting Cellular Network Hot Spots
  Using Sector Performance Indicators}. In \bibinfo{booktitle}{\emph{2017 IEEE
  33rd International Conference on Data Engineering (ICDE)}}.
  \bibinfo{pages}{259--270}.
\newblock
\urldef\tempurl%
\url{https://doi.org/10.1109/ICDE.2017.85}
\showDOI{\tempurl}


\bibitem[Smola and Sch{\"o}lkopf(2004)]%
        {Smola2004ATO}
\bibfield{author}{\bibinfo{person}{Alex Smola} {and} \bibinfo{person}{Bernhard
  Sch{\"o}lkopf}.} \bibinfo{year}{2004}\natexlab{}.
\newblock \showarticletitle{A tutorial on support vector regression}.
\newblock \bibinfo{journal}{\emph{Statistics and Computing}}
  \bibinfo{volume}{14} (\bibinfo{year}{2004}), \bibinfo{pages}{199--222}.
\newblock


\bibitem[Sun et~al\mbox{.}(2014)]%
        {SUN2014496}
\bibfield{author}{\bibinfo{person}{Huijun Sun}, \bibinfo{person}{Jianjun Wu},
  \bibinfo{person}{Dan Ma}, {and} \bibinfo{person}{Jiancheng Long}.}
  \bibinfo{year}{2014}\natexlab{}.
\newblock \showarticletitle{Spatial distribution complexities of traffic
  congestion and bottlenecks in different network topologies}.
\newblock \bibinfo{journal}{\emph{Applied Mathematical Modelling}}
  \bibinfo{volume}{38}, \bibinfo{number}{2} (\bibinfo{year}{2014}),
  \bibinfo{pages}{496--505}.
\newblock
\showISSN{0307-904X}
\urldef\tempurl%
\url{https://doi.org/10.1016/j.apm.2013.06.027}
\showDOI{\tempurl}


\bibitem[Sun et~al\mbox{.}(2017)]%
        {Sun2017}
\bibfield{author}{\bibinfo{person}{Yunchuan Sun}, \bibinfo{person}{Xinpei Yu},
  \bibinfo{person}{Rongfang Bie}, {and} \bibinfo{person}{Houbing Song}.}
  \bibinfo{year}{2017}\natexlab{}.
\newblock \showarticletitle{Discovering Time-Dependent Shortest Path on Traffic
  Graph for Drivers towards Green Driving}.
\newblock \bibinfo{journal}{\emph{J. Netw. Comput. Appl.}}
  \bibinfo{volume}{83}, \bibinfo{number}{C} (\bibinfo{date}{apr}
  \bibinfo{year}{2017}), \bibinfo{pages}{204–212}.
\newblock
\showISSN{1084-8045}
\urldef\tempurl%
\url{https://doi.org/10.1016/j.jnca.2015.10.018}
\showDOI{\tempurl}


\bibitem[Tang et~al\mbox{.}(2020)]%
        {Tang2020JointMO}
\bibfield{author}{\bibinfo{person}{Xianfeng Tang}, \bibinfo{person}{Huaxiu
  Yao}, \bibinfo{person}{Yiwei Sun}, \bibinfo{person}{Charu~C. Aggarwal},
  \bibinfo{person}{Prasenjit Mitra}, {and} \bibinfo{person}{Suhang Wang}.}
  \bibinfo{year}{2020}\natexlab{}.
\newblock \showarticletitle{Joint Modeling of Local and Global Temporal
  Dynamics for Multivariate Time Series Forecasting with Missing Values}.
\newblock \bibinfo{journal}{\emph{ArXiv}}  \bibinfo{volume}{abs/1911.10273}
  (\bibinfo{year}{2020}).
\newblock


\bibitem[Ting and Chai(2008)]%
        {ting2008}
\bibfield{author}{\bibinfo{person}{Y.~K. Ting, Stephanie} {and}
  \bibinfo{person}{T.~Tiong Chai}.} \bibinfo{year}{2008}\natexlab{}.
\newblock \showarticletitle{WCDMA Network planning and optimisation}.
  \bibinfo{pages}{317 -- 322}.
\newblock
\urldef\tempurl%
\url{https://doi.org/10.1109/NCTT.2008.4814295}
\showDOI{\tempurl}


\bibitem[Waldhauser et~al\mbox{.}(2012)]%
        {Waldhauser2012}
\bibfield{author}{\bibinfo{person}{Richard Waldhauser}, \bibinfo{person}{Markus
  Staufer}, \bibinfo{person}{Seppo Hamalainen}, \bibinfo{person}{Henning
  Sanneck}, \bibinfo{person}{Haitao Tang}, \bibinfo{person}{Lars Schmelz},
  \bibinfo{person}{Jürgen Goerge}, \bibinfo{person}{Paul Stephens},
  \bibinfo{person}{Krzysztof Kordybach}, {and} \bibinfo{person}{Clemens
  Suerbaum}.} \bibinfo{year}{2012}\natexlab{}.
\newblock \bibinfo{booktitle}{\emph{LTE Self-Organising Networks (SON): Network
  Management Automation for Operational Efficiency}}.
\newblock \bibinfo{pages}{39--80}.
\newblock
\showISBNx{9781119970675}
\urldef\tempurl%
\url{https://doi.org/10.1002/9781119961789.ch3}
\showDOI{\tempurl}


\bibitem[Winstein et~al\mbox{.}(2013)]%
        {Winstein2013}
\bibfield{author}{\bibinfo{person}{Keith Winstein}, \bibinfo{person}{Anirudh
  Sivaraman}, {and} \bibinfo{person}{Hari Balakrishnan}.}
  \bibinfo{year}{2013}\natexlab{}.
\newblock \showarticletitle{Stochastic Forecasts Achieve High Throughput and
  Low Delay over Cellular Networks}. In \bibinfo{booktitle}{\emph{Proceedings
  of the 10th USENIX Conference on Networked Systems Design and
  Implementation}} (Lombard, IL) \emph{(\bibinfo{series}{nsdi'13})}.
  \bibinfo{publisher}{USENIX Association}, \bibinfo{address}{USA},
  \bibinfo{pages}{459–472}.
\newblock


\bibitem[YANG et~al\mbox{.}(2019)]%
        {DiYANG20192018EDP7330}
\bibfield{author}{\bibinfo{person}{Di YANG}, \bibinfo{person}{Songjiang LI},
  \bibinfo{person}{Zhou PENG}, \bibinfo{person}{Peng WANG},
  \bibinfo{person}{Junhui WANG}, {and} \bibinfo{person}{Huamin YANG}.}
  \bibinfo{year}{2019}\natexlab{}.
\newblock \showarticletitle{MF-CNN: Traffic Flow Prediction Using Convolutional
  Neural Network and Multi-Features Fusion}.
\newblock \bibinfo{journal}{\emph{IEICE Transactions on Information and
  Systems}} \bibinfo{volume}{E102.D}, \bibinfo{number}{8}
  (\bibinfo{year}{2019}), \bibinfo{pages}{1526--1536}.
\newblock
\urldef\tempurl%
\url{https://doi.org/10.1587/transinf.2018EDP7330}
\showDOI{\tempurl}


\bibitem[Yu et~al\mbox{.}(2018)]%
        {Yu_2018}
\bibfield{author}{\bibinfo{person}{Bing Yu}, \bibinfo{person}{Haoteng Yin},
  {and} \bibinfo{person}{Zhanxing Zhu}.} \bibinfo{year}{2018}\natexlab{}.
\newblock \showarticletitle{Spatio-Temporal Graph Convolutional Networks: A
  Deep Learning Framework for Traffic Forecasting}. In
  \bibinfo{booktitle}{\emph{Proceedings of the Twenty-Seventh International
  Joint Conference on Artificial Intelligence}}.
  \bibinfo{publisher}{International Joint Conferences on Artificial
  Intelligence Organization}.
\newblock
\urldef\tempurl%
\url{https://doi.org/10.24963/ijcai.2018/505}
\showDOI{\tempurl}


\bibitem[Yu et~al\mbox{.}(2017)]%
        {Yu2017SpatiotemporalRC}
\bibfield{author}{\bibinfo{person}{Haiyang Yu}, \bibinfo{person}{Zhihai Wu},
  \bibinfo{person}{Shuqin Wang}, \bibinfo{person}{Yunpeng Wang}, {and}
  \bibinfo{person}{Xiaolei Ma}.} \bibinfo{year}{2017}\natexlab{}.
\newblock \showarticletitle{Spatiotemporal Recurrent Convolutional Networks for
  Traffic Prediction in Transportation Networks}.
\newblock \bibinfo{journal}{\emph{Sensors (Basel, Switzerland)}}
  \bibinfo{volume}{17} (\bibinfo{year}{2017}).
\newblock


\bibitem[Zhao et~al\mbox{.}(2020)]%
        {Zhao_2020}
\bibfield{author}{\bibinfo{person}{Ling Zhao}, \bibinfo{person}{Yujiao Song},
  \bibinfo{person}{Chao Zhang}, \bibinfo{person}{Yu Liu}, \bibinfo{person}{Pu
  Wang}, \bibinfo{person}{Tao Lin}, \bibinfo{person}{Min Deng}, {and}
  \bibinfo{person}{Haifeng Li}.} \bibinfo{year}{2020}\natexlab{}.
\newblock \showarticletitle{T-{GCN}: A Temporal Graph Convolutional Network for
  Traffic Prediction}.
\newblock \bibinfo{journal}{\emph{{IEEE} Transactions on Intelligent
  Transportation Systems}} \bibinfo{volume}{21}, \bibinfo{number}{9}
  (\bibinfo{date}{sep} \bibinfo{year}{2020}), \bibinfo{pages}{3848--3858}.
\newblock
\urldef\tempurl%
\url{https://doi.org/10.1109/tits.2019.2935152}
\showDOI{\tempurl}


\bibitem[Zhao et~al\mbox{.}(2017)]%
        {Zhao2017}
\bibfield{author}{\bibinfo{person}{Zheng Zhao}, \bibinfo{person}{Weihai Chen},
  \bibinfo{person}{Xingming Wu}, \bibinfo{person}{Peter C.~Y. Chen}, {and}
  \bibinfo{person}{Jingmeng Liu}.} \bibinfo{year}{2017}\natexlab{}.
\newblock \showarticletitle{LSTM network: a deep learning approach for
  short-term traffic forecast}.
\newblock \bibinfo{journal}{\emph{IET Intelligent Transport Systems}}
  \bibinfo{volume}{11}, \bibinfo{number}{2} (\bibinfo{year}{2017}),
  \bibinfo{pages}{68--75}.
\newblock
\urldef\tempurl%
\url{https://doi.org/10.1049/iet-its.2016.0208}
\showDOI{\tempurl}
\showeprint{https://ietresearch.onlinelibrary.wiley.com/doi/pdf/10.1049/iet-its.2016.0208}


\bibitem[Zhao et~al\mbox{.}(2021)]%
        {Zhao_2021}
\bibfield{author}{\bibinfo{person}{Zhen Zhao}, \bibinfo{person}{Ze Li},
  \bibinfo{person}{Fuxin Li}, {and} \bibinfo{person}{Yang Liu}.}
  \bibinfo{year}{2021}\natexlab{}.
\newblock \showarticletitle{CNN-LSTM Based Traffic Prediction Using
  Spatial-temporal Features}.
\newblock \bibinfo{journal}{\emph{Journal of Physics: Conference Series}}
  \bibinfo{volume}{2037}, \bibinfo{number}{1} (\bibinfo{date}{sep}
  \bibinfo{year}{2021}), \bibinfo{pages}{012065}.
\newblock
\urldef\tempurl%
\url{https://doi.org/10.1088/1742-6596/2037/1/012065}
\showDOI{\tempurl}


\bibitem[Zhou et~al\mbox{.}(2020)]%
        {LSTM_HOTSPOT}
\bibfield{author}{\bibinfo{person}{Lixia Zhou}, \bibinfo{person}{Xia Chen},
  \bibinfo{person}{Runsha Dong}, {and} \bibinfo{person}{Shan Yang}.}
  \bibinfo{year}{2020}\natexlab{}.
\newblock \showarticletitle{Hotspots Prediction Based on LSTM Neural Network
  for Cellular Networks}.
\newblock \bibinfo{journal}{\emph{Journal of Physics: Conference Series}}
  \bibinfo{volume}{1624} (\bibinfo{date}{10} \bibinfo{year}{2020}),
  \bibinfo{pages}{052016}.
\newblock
\urldef\tempurl%
\url{https://doi.org/10.1088/1742-6596/1624/5/052016}
\showDOI{\tempurl}


\bibitem[Zhuang and Cao(2022)]%
        {app12178714}
\bibfield{author}{\bibinfo{person}{Weiqing Zhuang} {and}
  \bibinfo{person}{Yongbo Cao}.} \bibinfo{year}{2022}\natexlab{}.
\newblock \showarticletitle{Short-Term Traffic Flow Prediction Based on
  CNN-BILSTM with Multicomponent Information}.
\newblock \bibinfo{journal}{\emph{Applied Sciences}} \bibinfo{volume}{12},
  \bibinfo{number}{17} (\bibinfo{year}{2022}).
\newblock
\showISSN{2076-3417}
\urldef\tempurl%
\url{https://doi.org/10.3390/app12178714}
\showDOI{\tempurl}


\end{thebibliography}

\end{document}